%% file: main.tex
\documentclass[fleqn,usenatbib]{mnras}

\usepackage{newtxtext,newtxmath}

\usepackage[T1]{fontenc}

\DeclareRobustCommand{\VAN}[3]{#2}
\let\VANthebibliography\thebibliography
\def\thebibliography{\DeclareRobustCommand{\VAN}[3]{##3}\VANthebibliography}

\usepackage{graphicx}	
\usepackage{amsmath}	
\usepackage{booktabs}   
\usepackage{siunitx}    
\usepackage{pgf}        
\usepackage{xspace}     
\usepackage{ifthen}     
\usepackage{enumitem}   
\usepackage{flushend}   
\usepackage[all]{hypcap} 
\usepackage[noabbrev,
            nameinlink,
            capitalise]{cleveref} 	 

\newcommand{\StoN}{\mathrm{S}/\mathrm{N}}
\newcommand\galaxyname[2]{#1\,#2}

\graphicspath{{fig/}}

\newcommand\ionized[2]{[\mathrm{#1}\,\textsc{#2}]}
\newcommand\OI[1][]{\ifthenelse{\equal{#1}{}}{\ionized{O}{i}}{\ionized{O}{i}\,\uplambda#1}}
\newcommand\OII[1][]{\ifthenelse{\equal{#1}{}}{\ionized{O}{ii}}{\ionized{O}{ii}\,\uplambda#1}}
\newcommand\OIII[1][]{\ifthenelse{\equal{#1}{}}{\ionized{O}{iii}}{\ionized{O}{iii}\,\uplambda#1}}
\newcommand\NII[1][]{\ifthenelse{\equal{#1}{}}{\ionized{N}{ii}}{\ionized{N}{ii}\,\uplambda#1}}
\newcommand\SII[1][]{\ifthenelse{\equal{#1}{}}{\ionized{S}{ii}}{\ionized{S}{ii}\,\uplambda#1}}
\newcommand\SIII[1][]{\ifthenelse{\equal{#1}{}}{\ionized{S}{iii}}{\ionized{S}{iii}\,\uplambda#1}}
\newcommand\HA[1][]{\ifthenelse{\equal{#1}{}}{\mathrm{H}\,\upalpha}{\mathrm{H}\,\upalpha\,\uplambda#1}}
\newcommand\HB[1][]{\ifthenelse{\equal{#1}{}}{\mathrm{H}\,\upbeta}{\mathrm{H}\,\upbeta\,\uplambda#1}}

\newcommand\RA[4]{$#1^\mathrm{h}#2^\mathrm{m}#3^\mathrm{s}\kern -3pt.#4$}
\newcommand\DEC[4]{$#1^\mathrm{d}#2^\mathrm{m}#3^\mathrm{s}\kern -3pt.#4$}

\newcommand\HII{H\,\textsc{ii}\xspace}

\newcommand\FUV{\mathrm{FUV}}
\newcommand\NUV{\mathrm{NUV}}
\newcommand\logOH{12+\log (\mathrm{O}/\mathrm{H})}
\newcommand\DeltaOH{\Delta (\mathrm{O}/\mathrm{H})}
\newcommand\EW{\mathrm{EW}(\HA)}

\DeclareSIUnit\angstrom{\text {Å}}
\DeclareSIUnit\percent{per\,cent}
\DeclareSIUnit\arcsec{arcsec}
\DeclareSIUnit\arcmin{arcmin}
\DeclareSIUnit\parsec{pc}
\DeclareSIUnit\lightyear{ly}
\DeclareSIUnit\year{yr}
\DeclareSIUnit\mag{mag}
\DeclareSIUnit\erg{erg}
\DeclareSIUnit\Msun{M_\odot}
\DeclareSIUnit\Lsun{L_\odot}

\newcommand*{\orcidlink}[1]{%
    \href{https://orcid.org/#1}{\,\raisebox{0.2em}{%
        \includegraphics[height=0.6em,width=0.6em]{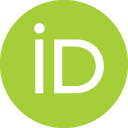}%
    }}}

\title[\HII region evolutionary sequence]{Stellar associations powering \HII regions -- I. Defining an evolutionary sequence}

\author[Scheuermann et al.]{Fabian Scheuermann\orcidlink{0000-0003-2707-4678},$^{\hyperlink{ARI}{1}}$\thanks{E-mail: f.scheuermann@uni-heidelberg.de} 
Kathryn Kreckel\orcidlink{0000-0001-6551-3091},$^{\hyperlink{ARI}{1}}$ 
Ashley T.~Barnes\orcidlink{0000-0003-0410-4504},$^{\hyperlink{Bonn}{2}}$ 
\newauthor
Francesco Belfiore\orcidlink{0000-0002-2545-5752},$^{\hyperlink{INAF}{3}}$ 
Brent Groves\orcidlink{0000-0002-9768-0246},$^{\hyperlink{Perth}{4}}$ 
Stephen Hannon\orcidlink{0000-0001-9628-8958},$^{\hyperlink{Riverside}{5},\hyperlink{Gemini}{6}}$ 
Janice C.~Lee\orcidlink{0000-0002-2278-9407},$^{\hyperlink{Gemini}{6},\hyperlink{Steward}{7}}$ 
\newauthor
Rebecca Minsley$^{\hyperlink{Arizona}{8}}$,
Erik Rosolowsky\orcidlink{0000-0002-5204-2259},$^{\hyperlink{Alberta}{9}}$ 
Frank Bigiel\orcidlink{0000-0003-0166-9745},$^{\hyperlink{Bonn}{2}}$ 
Guillermo A.~Blanc\orcidlink{0000-0003-4218-3944},$^{\hyperlink{Carnegie}{10},\hyperlink{Chile}{11}}$ 
\newauthor
M\'ed\'eric Boquien\orcidlink{0000-0003-0946-6176},$^{\hyperlink{Antofagasta}{12}}$ 
Daniel A. Dale\orcidlink{0000-0002-5782-9093},$^{\hyperlink{Wyoming}{13}}$ 
Sinan Deger\orcidlink{0000-0003-1943-723X},$^{\hyperlink{Stockholm}{14}}$ 
Oleg Egorov\orcidlink{0000-0002-4755-118X},$^{\hyperlink{ARI}{1}}$ 
\newauthor
Eric Emsellem\orcidlink{0000-0002-6155-7166},$^{\hyperlink{ESO}{15},\hyperlink{Lyon}{16}}$ 
Simon C.~O.~Glover\orcidlink{0000-0001-6708-1317},$^{\hyperlink{ITA}{17}}$ 
Kathryn Grasha\orcidlink{0000-0002-3247-5321},$^{\hyperlink{Canberra}{18},\hyperlink{ARC}{19}}$ 
Hamid Hassani\orcidlink{0000-0002-8806-6308},$^{\hyperlink{Alberta}{9}}$ 
\newauthor
Sarah M.~R.~Jeffreson\orcidlink{0000-0002-4232-0200},$^{\hyperlink{Harvard}{20}}$ 
Ralf S.~Klessen\orcidlink{0000-0002-0560-3172},$^{\hyperlink{ITA}{17},\hyperlink{IWR}{21}}$ 
J.~M.~Diederik~Kruijssen\orcidlink{0000-0002-8804-0212},$^{\hyperlink{cool}{22}}$ 
\newauthor
Kirsten L. Larson\orcidlink{0000-0003-3917-6460},$^{\hyperlink{STsci}{23}}$ 
Adam K. Leroy\orcidlink{0000-0002-2545-1700},$^{\hyperlink{Ohio}{24}}$ 
Laura A. Lopez\orcidlink{0000-0002-1790-3148},$^{\hyperlink{Ohio}{24},\hyperlink{Colombus}{25}}$ 
Hsi-An Pan\orcidlink{0000-0002-1370-6964},$^{\hyperlink{Taiwan}{26}}$ 
\newauthor
Patricia S\'anchez-Bl\'azquez\orcidlink{0000-0003-0651-0098},$^{\hyperlink{Cantoblanco}{27}}$ 
Francesco Santoro\orcidlink{0000-0002-6363-9851},$^{\hyperlink{MPIA}{28}}$ 
Eva Schinnerer\orcidlink{0000-0002-3933-7677},$^{\hyperlink{MPIA}{28}}$ 
\newauthor
David A. Thilker\orcidlink{0000-0002-8528-7340},$^{\hyperlink{JohnHopkins}{29}}$ 
Bradley C. Whitmore\orcidlink{0000-0002-3784-7032},$^{\hyperlink{Stsci}{23}}$ 
Elizabeth J. Watkins\orcidlink{0000-0002-7365-5791},$^{\hyperlink{ARI}{1}}$ 
\newauthor
Thomas G. Williams\orcidlink{0000-0002-0012-2142}\,$^{\hyperlink{MPIA}{28}, \hyperlink{Oxford}{30}}$ 
\vspace*{0.2em} \\
Affiliations are listed at the end of the paper.
}

\date{Accepted 2023 March 16. Received 2023 February 23; in original form 2022 November 5}

\pubyear{2023}

\begin{document}
\label{firstpage}
\pagerange{\pageref{firstpage}--\pageref{lastpage}}
\maketitle

\begin{abstract}
Connecting the gas in \HII regions to the underlying source of the ionizing radiation can help us constrain the physical processes of stellar feedback and how \HII regions evolve over time. 
With PHANGS--MUSE we detect nearly \num{24000} \HII regions across 19 galaxies and measure the physical properties of the ionized gas (e.g.\ metallicity, ionization parameter, density). 
We use catalogues of multi-scale stellar associations from PHANGS--\textit{HST} to obtain constraints on the age of the ionizing sources. 
We construct a matched catalogue of \num{4177} \HII regions that are clearly linked to a single ionizing association. 
A weak anti-correlation is observed between the association ages and the $\HA$ equivalent width $\EW$, the $\HA/\FUV$ flux ratio and the ionization parameter, $\log q$. 
As all three are expected to decrease as the stellar population ages, this could indicate that we observe an evolutionary sequence. 
This interpretation is further supported by correlations between all three properties. 
Interpreting these as evolutionary tracers, we find younger nebulae to be more attenuated by dust and closer to giant molecular clouds, in line with recent models of feedback-regulated star formation. 
We also observe strong correlations with the local metallicity variations and all three proposed age tracers, suggestive of star formation preferentially occurring in locations of locally enhanced metallicity. 
Overall, $\EW$ and $\log q$ show the most consistent trends and appear to be most reliable tracers for the age of an \HII region.
\end{abstract}





\input{"tex/01introduction"}
\input{"tex/02data"}
\input{"tex/03match_catalogues"}
\input{"tex/04evolutionary_sequence"}
\input{"tex/05conclusion"}

\section*{Acknowledgements}

We thank the anonymous referee for the helpful comments that improved this work. 
This work was carried out as part of the PHANGS collaboration. 
Based on observations collected at the European Southern Observatory under ESO programmes 094.C-0623 (PI: Kreckel), 095.C-0473,  098.C-0484 (PI: Blanc), 1100.B-0651 (PHANGS--MUSE; PI: Schinnerer), as well as 094.B-0321 (MAGNUM; PI: Marconi), 099.B-0242, 0100.B-0116, 098.B-0551 (MAD; PI: Carollo) and 097.B-0640 (TIMER; PI: Gadotti). 
Based on observations made with the NASA/ESA \textit{Hubble Space Telescope}, obtained from the data archive at the Space Telescope Science Institute. STScI is operated by the Association of Universities for Research in Astronomy, Inc. under NASA contract NAS 5-26555. Support for Program number 15654 was provided through a grant from the STScI under NASA contract NAS5-26555. This publication uses data from the \textit{AstroSat} mission of the Indian Space Research Organisation (ISRO), archived at the Indian Space Science Data Centre (ISSDC).
FS, KK and OE gratefully acknowledges funding from the German Research Foundation (DFG) in the form of an Emmy Noether Research Group (grant number KR4598/2-1, PI Kreckel).
KK and EW acknowledge support from the DFG via SFB 881 ‘The Milky Way System’ (project-ID 138713538; subproject P2).
ATB and FB would like to acknowledge funding from the European Research Council (ERC) under the European Union’s Horizon 2020 research and innovation programme (grant agreement No. 726384/Empire).
ER and HH acknowledges the support of the Natural Sciences and Engineering Research Council of Canada (NSERC), funding reference number RGPIN-2017-03987, and the Canadian Space Agency funding reference numbers SE-ASTROSAT19 and 22ASTALBER.
GAB acknowledges support from the ANID BASAL FB210003 project.
MB gratefully acknowledges support by the ANID BASAL project FB210003 and from the FONDECYT regular grant 1211000.
SD is supported by funding from the European Research Council (ERC) under the European Union’s Horizon 2020 research and innovation programme (grant agreement no. 101018897 CosmicExplorer).
RSK and SCOG acknowledge funding from the Deutsche Forschungsgemeinschaft (DFG) via SFB 881 ‘The Milky Way System’ (subprojects A1, B1, B2 and B8) and from the Heidelberg Cluster of Excellence STRUCTURES in the framework of Germany’s Excellence Strategy (grant EXC-2181/1-390900948). They also acknowledge support from the European Research Council in the ERC synergy grant ‘ECOGAL’ Understanding our Galactic ecosystem: From the disk of the Milky Way to the formation sites of stars and planets’ (project ID 855130).
SMRJ is supported by Harvard University through the ITC.
JMDK gratefully acknowledges funding from the European Research Council (ERC) under the European Union’s Horizon 2020 research and innovation programme via the ERC Starting Grant MUSTANG (grant agreement number 714907). COOL Research DAO is a Decentralised Autonomous Organisation supporting research in astrophysics aimed at uncovering our cosmic origins.
The work of AKL was partially supported by the National Science Foundation (NSF) under Grants No.~1653300 and 2205628.
HAP acknowledges support by the National Science and Technology Council of Taiwan under grant 110-2112-M-032-020-MY3.
ES and TGW acknowledges funding from the European Research Council (ERC) under the European Union’s Horizon 2020 research and innovation programme (grant agreement No. 694343).
This research made use of \textsc{astropy} \citep{Astropy+2013,Astropy+2018,Astropy+2022},  \textsc{numpy} \citep{Harris+2020}, \textsc{matplotlib} \citep{Hunter+2007} and \textsc{pyneb} \citep{Luridiana+2015}.
The distances in \cref{tbl:sample} were compiled by \citet{Anand+2021a} and are based on \citet{Freedman+2001,Nugent+2006,Jacobs+2009,Kourkchi+2017,Shaya+2017,Kourkchi+2020,Anand+2021a,Scheuermann+2022}.

\section*{Data availability}
The MUSE data underlying this article are presented in \citet{Emsellem+2022} and are available from the ESO archive\footnote{\url{https://archive.eso.org/scienceportal/home?data_collection=PHANGS}} and the CADC\footnote{\url{https://www.canfar.net/storage/vault/list/phangs/RELEASES/PHANGS-MUSE}}. 
The \textit{HST} data are presented in \citet{Lee+2022} and are available on the PHANGS--\textit{HST} website\footnote{\url{https://archive.stsci.edu/hlsp/phangs-hst}}. 
The catalogue with the background subtracted equivalent width and the matched catalogue with the \texttt{one-to-one sample} are both available in the online supplementary material of the journal. The code for this project can be found at
\begin{center}
\url{https://github.com/fschmnn/gas-around-stars}
\end{center}



\bibliographystyle{mnras}
\bibliography{paper} 



\appendix
\input{"tex/06appendix"}


\vspace{4mm}

\noindent {\itshape
\hypertarget{ARI}{$^{1}$Astronomisches Rechen-Institut, Zentrum f\"{u}r Astronomie der Universit\"{a}t Heidelberg, M\"{o}nchhofstra\ss e 12-14, 69120 Heidelberg, Germany} \\
\hypertarget{Bonn}{$^{2}$Argelander-Institut f\"{u}r Astronomie, Universit\"{a}t Bonn, Auf dem H\"{u}gel 71, D-53121 Bonn, Germany} \\
\hypertarget{INAF}{$^{3}$INAF -- Osservatorio Astrofisico di Arcetri, Largo E. Fermi 5, I-50157 Firenze, Italy} \\
\hypertarget{Perth}{$^{4}$International Centre for Radio Astronomy Research, University of Western Australia, 7 Fairway, Crawley, 6009 WA, Australia} \\
\hypertarget{Riverside}{$^{5}$Department of Physics and Astronomy, University of California, Riverside, CA 92501, USA} \\
\hypertarget{Gemini}{$^{6}$Gemini Observatory / NSF’s NOIRLab, 950 N. Cherry Avenue, Tucson, AZ 85719, USA} \\
\hypertarget{Steward}{$^{7}$Steward Observatory, University of Arizona, Tucson, AZ 85721, USA} \\
\hypertarget{Arizona}{$^{8}$Department of Physics and Astronomy, University of Arizona, USA} \\
\hypertarget{Alberta}{$^{9}$Department of Physics, University of Alberta, Edmonton, AB T6G2E1, Canada} \\
\hypertarget{Carnegie}{$^{10}$Observatories of the Carnegie Institution for Science, 813 Santa Barbara Street, Pasadena, CA 91101, USA} \\
\hypertarget{Chile}{$^{11}$Departamento de Astronom\'ia, Universidad de Chile, Camino del Observatorio 1515, Las Condes, Santiago, Chile} \\
\hypertarget{Antofagasta}{$^{12}$Centro de Astronomía (CITEVA), Universidad de Antofagasta, Avenida Angamos 601, Antofagasta, Chile} \\
\hypertarget{Wyoming}{$^{13}$Department of Physics \& Astronomy, University of Wyoming, Laramie, WY 82071, USA} \\
\hypertarget{Stockholm}{$^{14}$The Oskar Klein Centre for Cosmoparticle Physics, Department of Physics, Stockholm University, AlbaNova, Stockholm, SE-106 91, Sweden} \\
\hypertarget{ESO}{$^{15}$European Southern Observatory, Karl-Schwarzschild Stra\ss e 2, D-85748 Garching bei M\"{u}nchen, Germany} \\
\hypertarget{Lyon}{$^{16}$Univ Lyon, Univ Lyon 1, ENS de Lyon, CNRS, Centre de Recherche Astrophysique de Lyon UMR5574, F-69230 Saint-Genis-Laval, France} \\
\hypertarget{ITA}{$^{17}$Universit\"{a}t Heidelberg, Zentrum f\"{u}r Astronomie, Institut f\"{u}r Theoretische Astrophysik, Albert-Ueberle-Str. 2, 69120, Heidelberg, Germany} \\
\hypertarget{Canberra}{$^{18}$Research School of Astronomy and Astrophysics, Australian National University, Canberra, ACT 2611, Australia} \\
\hypertarget{ARC}{$^{19}$ARC Centre of Excellence for All Sky Astrophysics in 3 Dimensions (ASTRO 3D), Australia} \\
\hypertarget{Harvard}{$^{20}$Center for Astrophysics, Harvard \& Smithsonian, 60 Garden St, Cambridge, MA 02138, USA} \\
\hypertarget{IWR}{$^{21}$Universit\"{a}t Heidelberg, Interdisziplin\"{a}res Zentrum f\"{u}r Wissenschaftliches Rechnen, Im Neuenheimer Feld 205, D-69120 Heidelberg, Germany} \\
\hypertarget{cool}{$^{22}$Cosmic Origins Of Life (COOL) Research DAO, coolresearch.io} \\
\hypertarget{STsci}{$^{23}$Space Telescope Science Institute, 3700 San Martin Dr., Baltimore, MD 21218, USA} \\
\hypertarget{Ohio}{$^{24}$Department of Astronomy, The Ohio State University, 140 West 18th Avenue, Columbus, OH 43210, USA} \\
\hypertarget{Colombus}{$^{25}$Center for Cosmology and AstroParticle Physics, 191 West Woodruff Ave., Columbus, OH 43210, USA} \\
\hypertarget{Taiwan}{$^{26}$Department of Physics, Tamkang University, No.151, Yingzhuan Road, Tamsui District, New Taipei City 251301, Taiwan} \\
\hypertarget{Cantoblanco}{$^{27}$Departamento de F\'isica de la Tierra y Astrof\'isica, Universidad Complutense de Madrid, 28040, Madrid, Spain} \\
\hypertarget{MPIA}{$^{28}$Max Planck Institut für Astronomie, K\"{o}nigstuhl 17, 69117 Heidelberg, Germany} \\
\hypertarget{JohnHopkins}{$^{29}$Department of Physics and Astronomy, The Johns Hopkins University, Baltimore, MD 21218, USA} \\
\hypertarget{Oxford}{$^{30}$Sub-department of Astrophysics, Department of Physics, University of Oxford, Keble Road, Oxford OX1 3RH, UK} 
}

\bsp	
\label{lastpage} 
\end{document}

%% file: tex/01introduction.tex
\section{Introduction}\label{sec:introduction}

The formation of stars on galactic scales is a continuous cycle in which material from previous generations is recycled into new stars. 
This so-called \emph{baryon cycle} is regulated by the feedback from massive ($>\SI{8}{\Msun}$), short-lived stars \citep{Hopkins+2014,Kim+2017}. 
They produce ultraviolet (UV) radiation that ionizes the surrounding gas, forming \HII regions. Stellar winds and supernovae will further deposit energy in the cloud and enrich the gas, but the latter also mark the end of the short life \citep[\SIrange{3}{30}{\mega\year};][]{Ekstroem+2012} of the most massive O and B stars. 
If these feedback mechanisms are strong enough to overcome gravity, they are able to disperse the host cloud \citep{Dale+2014,Rahner+2017,Haid+2018,Kim+2018,Kruijssen+2019,Chevance+2022}, but if not, the \HII region will cease to exist once the last B stars explode. 
The exact ages of \HII regions are difficult to pin down, as line emission can vary as a function of multiple local physical conditions (age, but also e.g.\ metallicity or density). 
One approach is to determine the ages of underlying star clusters \citep[e.g.\ ][]{Whitmore+2011,Hollyhead+2015,Hannon+2019,Hannon+2022,Stevance+2020}, which can be estimated by fitting the observed \emph{spectral energy distribution} (SED) with theoretical models \citep[e.g.][]{Turner+2021}. 
Direct constraints from ionized nebulae on the other hand are rather rare, and existing studies are almost exclusively on $\si{\kilo\parsec}$ scales. One possibility is to use the $\HA$ \emph{equivalent width} $\EW$ \citep{Dottori+1981,Copetti+1986,Fernandes+2003,Levesque+2013} or the $\HA/\FUV$ ratio. 
Both fluxes are extensively used, most commonly as tracers for star formation \citep{Lee+2009} or to constrain the initial mass function \citep{Meurer+2009, Hermanowicz+2013}, but they have also been used on cloud scales as age indicators \citep[e.g.][]{SanchezGil+2011,Faesi+2014}.
$\HA$ is only created by the emission of the most massive stars in the stellar population. 
The stellar continuum in the $\FUV$ and underlying $\HA$, on the other hand, have a significant contribution from lower mass stars. 
Hence, once the most massive stars are gone, both $\EW$ and $\HA/\FUV$ start to decline. 
Assuming an instantaneous burst of star formation, population synthesis models like \textsc{starburst99} \citep{Leitherer+2014} or \textsc{bpass} \citep{Eldridge+2009}, the latter in combination with the photoionization model \textsc{cloudy} \citep{Ferland+2017}, also predict that after \SIrange{2}{3}{\mega\year} the ratio decreases monotonically with the age of the cluster.

Another potential age tracer is the \emph{ionization parameter} $q$, defined as the ratio of the incident ionizing photon flux $\Phi_\mathrm{H^0}$ to the local hydrogen density $n_\mathrm{H}$ \citep{Kewley+2002}. 
In the case of a spherical geometry, one can write the ionization parameter as
\begin{equation}
    q = \frac{\Phi_\mathrm{H^0}}{n_\mathrm{H}} = \frac{Q_\mathrm{H^0}}{4\pi R^2 n_\mathrm{H}},
\end{equation}  
where $Q_\mathrm{H^0}$ is the rate of ionizing photons striking the cloud at a distance $R$. Assuming the model of a partially-filled \emph{Strömgren sphere} \citep{Charlot+2001}, this becomes
\begin{equation}\label{eq:ionisation_stroemgren}
    q = \left( \alpha_\mathrm{B} \epsilon \right)^{2/3} \left( \frac{3\,Q_\mathrm{H^0}\, n_\mathrm{H}}{4\pi} \right)^{1/3},
\end{equation}
where $\alpha_\mathrm{B}$ is the case-B hydrogen recombination coefficient and $\epsilon$ is the volume-filling factor of the gas. 
As the stellar population ages, the rate at which it produces ionizing photons decreases \citep{Smith+2002} and so does the density as the \HII region expands (if the shell does not sweep up additional material), leading to a decrease of the ionization parameter over time. 
\citet{Dopita+2006} present a more realistic scenario and also conclude that the ionization parameter decreases as a function of time, with secondary dependencies on the ambient pressure, the cluster mass and the metallicity. 
Unlike the other two quantities, $\log q$ can not be measured directly, but the $\SIII/\SII$ ratio is a good proxy for it \citep{Diaz+1991}.

The physical conditions of the interstellar medium (ISM, e.g.\ metallicity, ionization parameter, density) regulate future star formation, and can be diagnosed via optical emission line ratios \citep{Kewley+2019b}. 
However some ratios can be sensitive to multiple properties, and careful work is needed to break those degeneracies  \citep{Kewley+2002,Kewley+2019b}. 
One particular issue is the relation between the chemical abundance and the ionization parameter. 
Some studies find an anti-correlation between the two \citep[e.g.][]{PerezMontero+2014,EspinosaPonce+2022}, while others find a positive correlation \citep[e.g.][]{Kreckel+2019,Grasha+2022}. 
This discrepancy is sometimes attributed to the resolution of the observations \citep{Kewley+2019b}, the type of galaxy that is studied \citep[star-forming or not,][]{Dopita+2014} or  to the underlying photoionization models that are used to measure the metallicity and ionization parameter \citep{Ji+2022}.

With an established age sequence, it becomes possible to directly track how conditions in the ISM evolve as a result of stellar feedback processes. 
However, it is challenging to identify an ionizing source for each \HII region and the sample for which we are able to is therefore biased. 
Direct age constraints from the ionized gas would allow us to include \HII regions without an ionizing source and hence study an unbiased populations.
With this, it becomes possible to study the evolutionary timescales, using large representative samples of \HII regions to probe the different stages in their evolution.
Modern integral field unit (IFU) spectroscopic surveys enable us to isolate individual \HII regions at $<\SI{100}{\parsec}$ scale and observe thousands of \HII regions across individual galaxies. 
Previous work has typically focused on detailed case studies of individual galaxies \citep{Niederhofer+2016,McLeod+2020,McLeod+2021,DellaBruna+2021}. 
In order to investigate the dependence on galactic properties like environments (morphological features), metallicities or star formation rates, a more comprehensive sample is required. 
The Physics at High Angular resolution in Nearby GalaxieS (PHANGS)\footnote{\url{http://www.phangs.org}} collaboration studies star formation in nearby galaxies, using large samples of giant molecular clouds (GMCs), \HII regions and star clusters. 
It combines observations from ALMA \citep{Leroy+2021b}, MUSE \citep{Emsellem+2022}, \textit{HST} \citep{Lee+2022} and \textit{JWST} \citep{Lee+2023} for a sample of 19 nearby galaxies.
This provides an unprecedented sample that allows us to study the ionized gas in \HII regions together with the massive stars that ionize them. 
We produce for the first time a catalogue of cross-matched ionizing sources and ionized nebulae, well suited for addressing how  stellar feedback evolves (empirically) and constraining stellar feedback models.

This paper is organised as follows: in \cref{sec:data}, we present the data and the existing catalogues that are used in the analysis. 
In \cref{sec:matched_catalogue} we match the \HII regions to their ionizing sources. 
In \cref{sec:evolutionary_sequence} we establish the \HII region evolutionary sequence and discuss our findings and conclude in \cref{sec:conclusion}.

%% file: tex/02data.tex
\section{Data}\label{sec:data}

\input{tbl/sample}

We perform a joint analysis of the 19 PHANGS galaxies listed in \cref{tbl:sample}, combining MUSE optical spectroscopy, \textit{HST} stellar association catalogues, and \textit{AstroSat} $\FUV$ imaging.
The galaxies in the sample have masses in the range of $9.4<\log(M/\si{\Msun})<11.0$ and star formation rates  $-0.56<\log(\mathrm{SFR}/\si{\Msun\per\year})<1.23$. They contain diverse morphological features like bars, rings, and active galactic nuclei, enabling us to study the impact of a wide variety of parameters.

\subsection[MUSE HII region catalogue]{MUSE \HII region catalogue}

To trace the ionized gas, we use IFU data observed by the Multi Unit Spectroscopic Explorer \citep[MUSE,][]{Bacon+2010} at the Very Large Telescope (VLT) in Chile. 
The PHANGS--MUSE survey (PI: Schinnerer) and the data reduction are described in \citet{Emsellem+2022}. 
They produced reduced and mosaicked spectral cubes and additional high-level data products like emission line maps. 
These maps achieve an average resolution of $\SI{0.72}{\arcsec}$, corresponding to a spatial resolution between \SIrange[range-phrase = {\text{~and~}}]{18.4}{77.9}{\parsec}, depending on the distance to the galaxy.

\citet{Santoro+2022} and \citet{Groves+2023} utilised these products to create a catalogue of \HII regions, which we use in this work and describe briefly below. 
\textsc{hiiphot} \citep{Thilker+2000} was used on the $\HA$ line maps to define the boundaries of the \HII regions based on their surface brightness profile. 
The integrated spectrum of each region was extracted from the spectral cubes and a number of nebular emission lines were fitted. 
We do not correct for the contribution of the diffuse ionized gas (DIG), as it is very sensitive to the exact \HII region boundaries and the majority of our \HII regions are bright ($L(\HA) >\SI{e37}{\erg\per\second}$ and $\HA$ surface brightness $>\SI{e39}{\erg\per\second\per\kilo\parsec\squared}$).
The fluxes were then corrected for Milky Way extinction with the extinction curve from \citet{ODonnell+1994}, adapting $R_V=3.1$ and $E(B-V)$ from \citet{Schlafly+2011}. For the internal extinction, the colour excess $E(B-V)$ is measured from the \emph{Balmer decrement}, assuming a theoretical ratio of $\HA/\HB=2.86$, and all fluxes are corrected assuming the same extinction curve parameters.

To remove contaminants such as supernova remnants or planetary nebulae, the \emph{Baldwin-Phillips-Terlevich} (BPT) diagram \citep{Baldwin+1981} is used to classify the objects \citep[based on the demarcation lines by][using the $\NII$, $\SII$ and $\OI$ lines]{Kauffmann+2003,Kewley+2001b}. 
For each line used, we require a $\StoN>5$ to be included in the final catalogue. This conservative $\StoN$ cut accounts for the possibility that the line flux uncertainties are slightly underestimated \citep{Emsellem+2022}. Across 19 galaxies, a total of \num{31497} nebulae are detected and \num{23736} of them are classified as \HII regions.

Next, a number of physical properties are derived. 
We calculate $\EW$ following the procedure outlined in \citet{Westfall+2019}.
The $\HA$ and continuum flux are obtained by directly summing the spectral bins in the data cube.
Our \HII regions are observed in projection against the stellar disk, which is dominated by light from older stars that are not physically associated with the massive star-forming regions. 
To correct for this contribution to our $\EW$ measurement, we estimate the background contributed by the stellar continuum from a spectrum integrated over an annular mask. 
These masks are created by growing the existing nebula masks until they have increased by a factor of three in area, excluding any pixels that fall inside of other nebulae. 
We then calculate the background corrected equivalent width, $\EW_\mathrm{corr}$, by subtracting the background from continuum (but not from the $\HA$ line flux, see \cref{sec:bkg_EW} for more details).

The gas phase abundance $\logOH$ is measured with the $S$-calibration from \citet{Pilyugin+2016} and the radial abundance gradient is fitted and subtracted from the individual nebulae to get the local metallicity offset $\DeltaOH$ \citep{Kreckel+2019}. 
The ionization parameter $q=U/c$ is derived from the $\SIII/\SII$ ratio, based on the calibration by \citet{Diaz+1991}:
\begin{equation}\label{eq:logq_D91}
        \log U=-1.684\cdot \log(\SIII/\SII)-2.986,
\end{equation}
where $\SIII/\SII=\SIII[(9069+9532)]\,/\,\SII[(6717+6731)]$. 
We measure $\SIII[9069]$ directly and assume a fixed atomic ratio of $\SIII[9532] = 2.5\cdot\SIII[9069]$ \citep{Osterbrock+2006}.

Finally we use \textsc{pyneb} \citep{Luridiana+2015} to derive the electron density from the $\SII[6731/6717]$ ratio. 
Ideally one would do this in conjunction with the temperature (e.g.\ from the $\NII[5755/6548]$ ratio). 
However, only $\sim\kern-2pt\SI{4}{\percent}$ of the \HII regions have a strong enough detection ($\StoN>10$) in the $\NII[5755]$ line to measure a temperature. 
We therefore assume a constant temperature of $\SI{8000}{\kelvin}$ for all nebulae. This is close to the mean temperature that we measure for the sub-sample that has a detection in the $\NII[5755]$ line \citep{Kreckel+2022}. 
That being said, the choice of $T_\mathrm{e}$ has only minor effects on the derived density (less than $\SI{10}{\percent}$, see also \cref{sec:CrossTemDen} for a more careful analysis).
More important is that most of the \HII regions are close to the low-density limit.
We apply the procedure outlined in \citet{Barnes+2021} and do not report densities for the regions that have $\SII[6731/6717]$ ratios within $3\sigma$ of this limit, as we are unable to derive reliable density measurements for them.

\subsection{\textit{HST} Stellar association catalogue}

The resolution of ground-based observations ($\sim\kern-5pt\SI{1}{\arcsec}$, corresponding to an average spatial resolution of $\sim\kern-2pt\SI{70}{\parsec}$ for the galaxies in our sample) is not sufficient to resolve the star clusters and associations that are the origin of the ionizing radiation and only the \textit{Hubble Space Telescope (HST)} is able to resolve the required $\sim\kern-4pt\SI{10}{\parsec}$ spatial scales \citep{Ryon+2017} in the optical and UV. 
The PHANGS--\textit{HST} survey (PI: Lee) covers 38 galaxies, including all galaxies in the PHANGS--MUSE sample, and is described in \citet{Lee+2022}. 
The galaxies were observed in 5 bands: F275W ($\NUV$), F336W (\textit{U}), F438W (\textit{B}), F555W (\textit{V}), F814W (\textit{I}). 

A number of high-level data products were produced by the PHANGS--\textit{HST} team. 
First is a catalogue of compact star clusters \citep[we use IR3 in this work]{Thilker+2022}. 
They are identified as sources that are slight more extended than a point source, based on their concentration index. 
However, the overlap with the \HII region catalogue is rather limited (as expected, as most clusters have ages $>\SI{10}{Myr}$) and we find that the majority of the ionizing sources are in OB associations \citep{Goddard+2010,Kruijssen+2012,Adamo+2015,Adamo+2020}. 
These larger and loosely bound structures were identified by \citet{Larson+2023} at different spatial scales and we use their multi-scale stellar association catalogue for our analysis.
A summary of the procedure is as follows, and also provided in \citet{Lee+2022}.
First, \textsc{dolphot} is used to create a catalogue of point-like sources (using either the $\NUV$ or \textit{V}-band). 
Then a `tracer image' is generated by smoothing to a fixed resolution (either $\SI{8}{\parsec}$, $\SI{16}{\parsec}$, $\SI{32}{\parsec}$ or $\SI{64}{\parsec}$) and a local background correction is applied to each by subtracting the equivalent image generated at four times larger scale. 
Next a \emph{watershed algorithm} \citep{Beucher+1979,vanderWalt+2014} is used to define the boundaries of the stellar associations that are identified at each scale as local enhancements in the background-subtracted tracer image.
Fluxes for the five available \textit{HST} filters are measured, using the photometry of detected stars, contained within the association.
Age, mass and reddening are derived by fitting theoretical models of a single stellar population \citep{Bruzual+2003} to the observed SED with \textsc{cigale} \citep{Boquien+2019}, assuming a fully sampled \citet{Chabrier+2003} IMF. The gas-phase metallicities are close to solar for most galaxies \citep{Groves+2023}, suggesting that the assumption of solar  metallicity is reasonable for the young stellar populations probed here. Complete details on the SED fitting are provided in \cite{Turner+2021}.

For this work, we choose to use the $\SI{32}{\parsec}$ stellar associations identified from the $\NUV$ images. 
This catalogue contains \num{16295} associations that fall within the MUSE field of view (FoV). 
This was done as the $\NUV$ is better at tracing the young massive stars that we are interested in, and the $\SI{32}{\parsec}$ structures are well matched to our MUSE resolution. 
While we find a larger number of \HII regions that overlap with the $\SI{64}{\parsec}$ associations, the percentage with a one-to-one relation is smaller, especially for the more nearby galaxies.
It should be noted that using a different scale for the analysis does not impact the results significantly.

\subsection{\textit{AstroSat}}

We used the Indian space telescope \textit{AstroSat} \citep{Singh+2014} to observe our galaxy sample in the far ultraviolet ($\FUV$, PI: Rosolowsky). 
The sample and data reduction are described in Hassani et al.\ (in preparation). The images have a resolution of $\SI{1.4}{\arcsec}$ with a pixel size of $\SI{0.42}{\arcsec}$, reasonably well matched to our MUSE resolution. 
Observations are available for 16 of the 19 MUSE galaxies (\galaxyname{NGC}{1087} and \galaxyname{NGC}{4303} can not be observed and for \galaxyname{NGC}{5068}, observing time is awarded but not yet executed). 
When available, we use the F148W filter, which is the case for all galaxies but \galaxyname{NGC}{1433}, \galaxyname{NGC}{1512} and \galaxyname{NGC}{4321}, which were observed with the F154W filter.
We subtract a uniform foreground that originates from Milky Way and solar system sources and whose average contribution varies between galaxies from \SIrange{1}{9}{\percent}.
The \textit{AstroSat} images are then reprojected to the MUSE images to measure the $\FUV$ fluxes in the spatial masks of the nebula catalogue. 
The fluxes are corrected for Milky Way and internal extinction, following the procedure described for the \HII regions. 
For the colour excess we use the value derived from the Balmer decrement and assume $E(B-V)_\mathrm{stellar} = X \cdot E(B-V)_\mathrm{Balmer}$. 
It is common to use $X=0.44$ to account for the difference between the colour excess derived from the Balmer decrement and the stellar continuum \citep{Calzetti+2000}, reflecting that nebular sources are often more closely associated with dusty star-forming complexes in contrast to the bulk stellar light \citep{Charlot+2000}.
However, we are preferentially targeting stellar continuum emission that we believe to be co-spatial with the nebular emission, and so instead assume $X=1.0$. 
In \cref{sec:validate} we compare the colour excess derived from the Balmer decrement to the one from the SED fit and find good agreement between the two, validating this choice.

%% file: tbl/sample.tex
\begin{table}
\centering
\caption{Galaxy sample with number of \HII regions \citep[from][]{Santoro+2022,Groves+2023} and stellar associations \citep[from][]{Larson+2023}. This only includes \HII regions and associations that are inside the overlapping FoV of \textit{HST} and MUSE. $N_\mathrm{match}$ is the number of objects in the \texttt{one-to-one sample} (see \cref{sec:matched_catalogue}).}
\begin{tabular}{r
                S[table-format=2.2]@{\,\( \pm \)\,}
                S[table-format=1.2]
                S[table-format=2.1]
                S[table-format=5]
                S[table-format=5]
                S[table-format=4]
                }
\toprule
Name & \multicolumn{2}{c}{distance$^\mathrm{a}$} & \multicolumn{1}{c}{resolution$^\mathrm{b}$}  & {$N_\mathrm{\HII}$} & {$N_\mathrm{asc}$} & {$N_\mathrm{match}^\mathrm{c}$}  \\
 & \multicolumn{2}{c}{$(\si{\mega\parsec})$} & \multicolumn{1}{c}{$(\si{\parsec})$}&  &  &   \\
\midrule
\galaxyname{IC}{5332}  &  9.01 & 0.41 & 31.4 & 608 & 397 & 135  \\
\galaxyname{NGC}{0628} &  9.84 & 0.63 & 34.7 & 1651 & 1141 & 379  \\
\galaxyname{NGC}{1087} & 15.85 & 2.24 & 56.9 & 892 & 487 & 181  \\
\galaxyname{NGC}{1300} & 18.99 & 2.85 & 57.9 & 1147 & 401 & 179 \\
\galaxyname{NGC}{1365} & 19.57 & 0.78 & 77.9 & 445 & 489 & 90 \\
\galaxyname{NGC}{1385} & 17.22 & 2.58 & 40.9 & 922 & 525 & 131 \\
\galaxyname{NGC}{1433} & 18.63 & 1.86 & 58.7 & 729 & 494 & 223  \\
\galaxyname{NGC}{1512} & 18.83 & 1.88 & 72.9 & 479 & 521 & 197 \\
\galaxyname{NGC}{1566} & 17.69 & 2.00 & 54.9 & 1448 & 1582 & 297 \\
\galaxyname{NGC}{1672} & 19.40 & 2.91 & 67.7 & 1047 & 1247 & 237 \\
\galaxyname{NGC}{2835} & 12.22 & 0.94 & 50.5 & 777 & 649 & 192 \\
\galaxyname{NGC}{3351} &  9.96 & 0.33 & 35.7 & 769 & 708 & 249 \\
\galaxyname{NGC}{3627} & 11.32 & 0.48 & 42.3 & 1016 & 1325 & 183 \\
\galaxyname{NGC}{4254} & 13.10 & 2.80 & 36.9 & 2375 & 1661 & 390 \\
\galaxyname{NGC}{4303} & 16.99 & 3.04 & 47.8 & 1956 & 1736 & 342  \\
\galaxyname{NGC}{4321} & 15.21 & 0.49 & 47.2 & 984 & 1483 & 332 \\
\galaxyname{NGC}{4535} & 15.77 & 0.37 & 33.7 & 1168 & 469 & 175 \\
\galaxyname{NGC}{5068} &  5.20 & 0.21 & 18.4 & 1405 & 715 & 174  \\
\galaxyname{NGC}{7496} & 18.72 & 2.81 & 71.6 & 523 & 265 & 91 \\\midrule
Total & \multicolumn{2}{c}{ } & & 20341 & 16295 & 4177  \\
\bottomrule\addlinespace
\multicolumn{7}{l}{$^\mathrm{a}$ From \citet{Anand+2021a}, see also the acknowledgements.}\\
\multicolumn{7}{l}{$^\mathrm{b}$ Based on the average value of all MUSE pointings in that galaxy.} 
\end{tabular}
\label{tbl:sample}
\end{table}

%% file: tex/03match_catalogues.tex
\section{Matching nebulae and stellar associations}
\label{sec:matched_catalogue}

\begin{figure}
\centering
\includegraphics[width=\columnwidth]{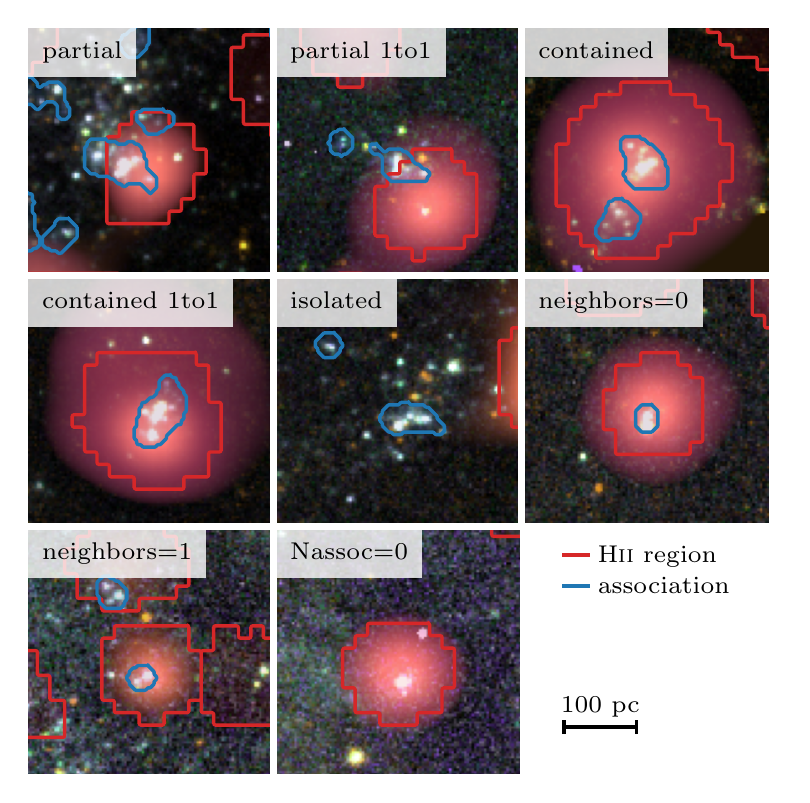}
\caption{Examples for the overlap between the \HII regions and stellar associations in \galaxyname{NGC}{1365}. The cutouts show a three colour composite images, based on the 5 available \textit{HST} bands, overlaid with the $\HA$ line emission of MUSE in red. The boundaries of the \HII regions are shown in red and the stellar associations in blue. Flags that characterise the overlap between the two catalogues are showcased in this figure.} 
\label{fig:overlap}
\end{figure}

To relate the \HII regions to the source of the ionizing radiation, we match the nebula catalogue with the association catalogue. 
Both have spatial masks, but with different pixel scales ($\SI{0.2}{\arcsec}$ per pixel for MUSE and $\SI{0.04}{\arcsec}$ per pixel for \textit{HST}). 
The coarser resolution of the nebula masks can result in some larger \HII regions that are not correctly partitioned \citep[see][]{Barnes+2022}. 
However we would still correctly map them to their ionizing sources.
The absolute astrometric accuracy is excellent for both data sets. 
\textit{HST} achieves $0.1$ pixel accuracy while MUSE reaches $0.5$ pixel, corresponding to two and a half \textit{HST} pixels.
Considering the uncertainties in deriving the exact boundaries of the \HII regions, this is negligible when matching the datasets.
We reproject the nebula masks to the association masks and determine the fractional overlap as measured in terms of the association or nebula pixel areas. 
In \cref{fig:overlap} we showcase examples for the overlap between the \HII regions and associations. 
For each of the stellar associations we define the \texttt{overlap} to be one of the following three:
\begin{itemize}[leftmargin=*]
    \item \texttt{isolated}: the association does not overlap with any nebula (fractional overlap of $\SI{0}{\percent}$). 
    \item \texttt{partial}: part of the association overlaps with one or more nebula, but part of it extends beyond it (fractional overlap between $0$ and $\SI{100}{\percent}$). 
    \item \texttt{contained}: the association is fully contained in the nebulae. It can be contained within multiple nebulae (fractional overlap of $\SI{100}{\percent}$).
 \end{itemize}
 For the \HII regions we determine the following properties:
 \begin{itemize}[leftmargin=*]
    \item \texttt{neighbors}: number of neighbouring \HII regions (i.e.\ regions that share a common boundary)
    \item \texttt{Nassoc}: number of stellar associations that overlap with the \HII region. 
\end{itemize}
Finally, we construct a catalogue of matched objects where each \HII region and stellar association overlap with exactly one stellar association and \HII region respectively. 
It contains \num{4177} objects that constitute the parent sample that we will be working with and to which we refer to as the \texttt{one-to-one sample}. This catalogue is available in the supplementary online material and incorporates a number of columns from the nebula and association catalogues. \Cref{tbl:matched_catalogue_columns} gives an overview of the properties that are included.

\input{tbl/matched_catalogue_columns}

\subsection{Statistics of the matched catalogue}

\input{tbl/overlap_stat}

From the initial nebula catalogue, \num{20341} \HII regions fall inside the \textit{HST} FoV and \num{8654} of them ($\SI{42.5}{\percent}$) overlap with at least one stellar association. 
Conversely, the association catalogue contains \num{16295} associations in the MUSE FoV and \num{11699} of them ($\SI{71.8}{\percent}$) overlap with at least one \HII region. 
\cref{tbl:overlap_statistics} provides a detailed breakdown of how the \HII regions and associations overlap. 

In cases where a nebula contains multiple associations, it is difficult to assess the contribution of the individual objects and ambiguous to assign a single age, as they are not necessarily coeval \citep{Efremov+1998}. 
In fact, when we find multiple stellar associations in one \HII region, they rarely have the same age. 
The age spread is usually small ($<\SI{10}{\mega\year}$), but there are also more extreme cases, with $\sim\kern-2pt\SI{25}{\percent}$ having larger age spreads.
Therefore, to simplify our analysis, in this paper we require a one-to-one match (this can be with either \texttt{partial} or \texttt{contained} overlap) between the associations and \HII regions, leaving us with a sample of \num{4177} objects in our parent \texttt{one-to-one sample} (see \cref{tbl:sample} for a detail breakdown by galaxy). 
Depending on the application, we also apply further cuts in age, mass or overlap to obtain cleaner samples. 

The nebula masks are dense and cover most of the spiral arms and hence it is likely that some of the overlaps are coincidental. 
To estimate how many of our matches this might concern, we run a test where we rotate the full set of nebula masks by $\SI{90}{\deg}$ around the galaxy centre, before matching with the associations. 
While previously one third of the \HII regions overlapped with an association, this number drops to one sixth. 
The number of \texttt{partial} overlapped associations decreases only slightly to $\SI{81}{\percent}$. 
The difference is more apparent with the \texttt{contained} sample: we estimate that only around $\SI{16}{\percent}$ of the \texttt{contained} associations could be chance alignments.
This motivates us to only use the \texttt{contained} objects for our analysis, reducing the sample to \num{1918} \HII regions and stellar associations.

\begin{figure}
    \centering
    \includegraphics{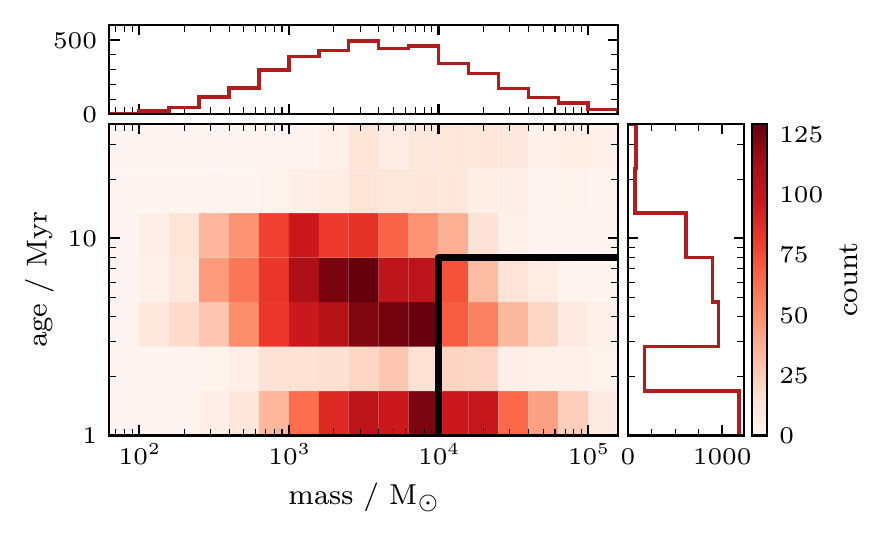}
    \caption{Distribution of masses and ages of the stellar associations in the matched catalogue (the \texttt{one-to-one} sample). The black lines mark the cuts that we apply to the sample to ensure a fully sampled IMF (more massive than $>\SI{e4}{\Msun}$) and to only include young clusters that should be associated with ionized gas (younger than $\leq\SI{8}{\mega\year}$). The mass cut leaves us with a sample of \num{1014} objects and the age cut with \num{3531}. Applying both cuts results in a sample of \num{756} objects.}
    \label{fig:catalogue_properties_2D_hist_v2}
\end{figure}

Another concern is stochastic sampling of the IMF \citep{Fouesneau+2010,Hannon+2019}. 
Below a certain cluster mass, the presence or absence of a single massive star can have a significant impact on the observed colours and hence the derived properties of the association, as well as on whether it is able to ionize hydrogen and has an associated detectable \HII region. 
We assume that stellar associations that are more massive than $\SI{e4}{\Msun}$ will fully sample the IMF \citep{daSilva+2012}.
We also do not expect significant $\HA$ emission to be associated with old stellar populations and therefore introduce an age cut at $\SI{8}{\mega\year}$.
The distribution of masses and ages of the matched catalogue is shown in \cref{fig:catalogue_properties_2D_hist_v2}, with both cuts illustrated by black lines. 
Note that the gap at \SIrange{2}{3}{\mega\year} is inherited from the stellar association catalogue, and reflects specific features in the stellar population tracks \citep{Larson+2023}. 
In total, \num{1041} objects pass the mass cut and \num{3531} pass the age cut. 
Applying both of these criteria, and further requiring the stellar associations to be \texttt{contained}, leaves \num{469} objects, constituting our \texttt{robust sample}.

\subsection[The physical nature of HII regions without associations]{The physical nature of \HII regions without associations}
\label{sec:unmatched_HII_regions}

\begin{figure}
    \centering
    \includegraphics{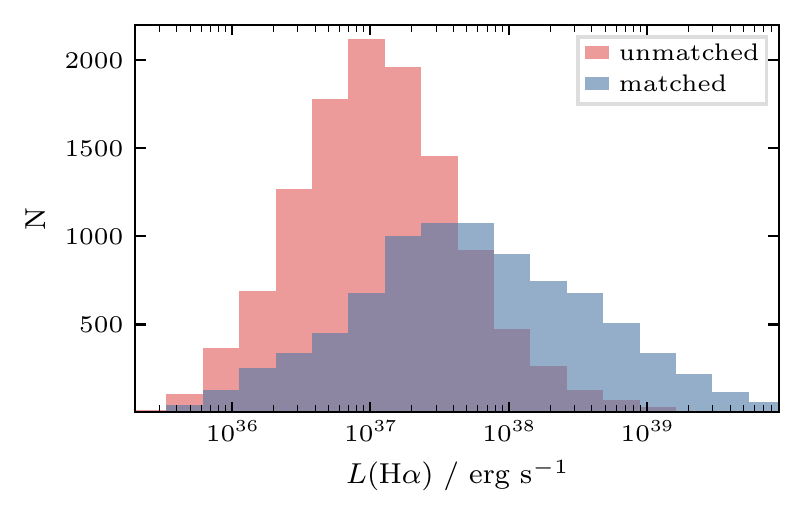}
    \caption{$\HA$ luminosity function of the \HII regions that overlap with one or more associations and those that do not. The matched \HII regions are on average a factor 10 brighter than the unmatched ones.}
    \label{fig:luminosity_function}
\end{figure}


As shown in \cref{tbl:overlap_statistics}, the majority of our \HII regions do not overlap with an association, which raises the question of the origin of their ionizing radiation. 
We consider several scenarios to explain this discrepancy. 
First of all, the unmatched \HII regions could host deeply embedded, highly extincted stars that we are unable to observe.
However their reddening distribution is almost identical to the matched \HII regions with a mean of $E(B-V)_\mathrm{Balmer}=0.26$, making it unlikely that we miss many objects due to extinction.

Secondly, we restricted our analysis to the $\SI{32}{\parsec}$ scale associations, but there are some variations between the different scales and $\SI{20.6}{\percent}$ of the unmatched \HII regions overlap with an $\SI{16}{\parsec}$ or $\SI{64}{\parsec}$ association. By design, the stellar association catalogue only includes objects with multiple peaks, excluding other ionizing sources like compact star clusters or individual stars. 
Looking at the $\HA$ luminosity function in \cref{fig:luminosity_function}, we see that \HII regions that are matched to an association are on average brighter by a factor of 10 compared to those that are unmatched, meaning that the unmatched regions are ionized by fainter sources.
We consider the previously mentioned PHANGS--\textit{HST} compact cluster by \citep[for example, the unmatched \HII region in \cref{fig:overlap} contains an object that is classified as a compact cluster]{Thilker+2022}. 
Using the machine learning classified catalogue \citetext{\citealp{Wei+2020}; Hannon et al.\ in preparation}, we find that \num{2569} of all \HII regions contain a compact cluster ($\SI{15.3}{\percent}$, comparable to the $\SI{13}{\percent}$ reported by \citealt{DellaBruna+2022b} for \galaxyname{M}{83}). 
The majority of these clusters are already contained in an association and only $\SI{6.1}{\percent}$ of the previously unmatched \HII regions contain a compact cluster. 
By design, the compact cluster identification pipeline disfavours the selection of looser association-like structures that are young. 
Considering that we are primarily interested in young and massive clusters, the number of potentially interesting compact clusters is only of the order of a few dozen.

Even after accounting for different scale associations and compact clusters, $\SI{43.5}{\percent}$ of the \HII regions are still without a catalogued ionizing source. However, looking at the \textit{HST} $\NUV$ images, we find peaks that are clearly associated with those unmatched \HII regions. 
Their absence in the aforementioned catalogues can have a few reasons. 
First, the selection of cluster candidates is based on the \textit{V} band and hence does not necessarily trace the ionizing sources as well as the $\NUV$. 
Secondly, they are often too compact to be classified as a star cluster according to their concentration index. 
As for the association catalogue, multiple peaks are required and the objects here are mostly single peaks. 
Using the point source catalogue, created with \textsc{dolphot}, that forms the base for both the stellar association and compact cluster catalogues \citep{Thilker+2022}, we find peaks in most \HII regions. 
If we require that a peak is detected with a $\StoN>5$ in both the $\NUV$ and at least one other filter, around $\SI{63.7}{\percent}$ of the unmatched \HII regions contain a peak. 

The fainter unmatched \HII regions ($L_{\mathrm{H}\alpha}\lesssim\SI{e37}{\erg\per\second}$) fall in a regime where they could be ionized by a single O-star \citep{Martins+2005}. 
The apparent magnitudes of the \textsc{dolphot} sources inside those unmatched \HII regions are also comparable to the luminosity of a single O-star (see also \cref{sec:single_stars} with \cref{fig:dolphot_peaks_stellar_models} for a comparison with models). 
This subsample of \HII regions that are potentially ionized by a single star (or a very small star cluster, dominated by a single massive star) is interesting in its own right, however beyond the scope of this paper. 

For now, we leave this with the understanding that most of the unmatched \HII regions in our catalogue have an ionizing source that is visible in the \textit{HST} data, but that these sources do not meet the criteria for our catalogue of stellar associations and hence are not included in our analysis.
Overall, we are able to identify $\NUV$-bright ionizing sources for the vast majority of our sample. 
After accounting for the different sources of ionizing radiation, only $\SI{18.6}{\percent}$ of the \HII regions are left without an ionizing counterpart. 
This number could be even further decreased by lowering the S/N requirement for the \textsc{dolphot} peaks (to $\SI{9.9}{\percent}$ with $\StoN>3$).

\subsection{Validating the sample}
\label{sec:validate}

\begin{figure}
    \centering
    \includegraphics[width=0.9\columnwidth]{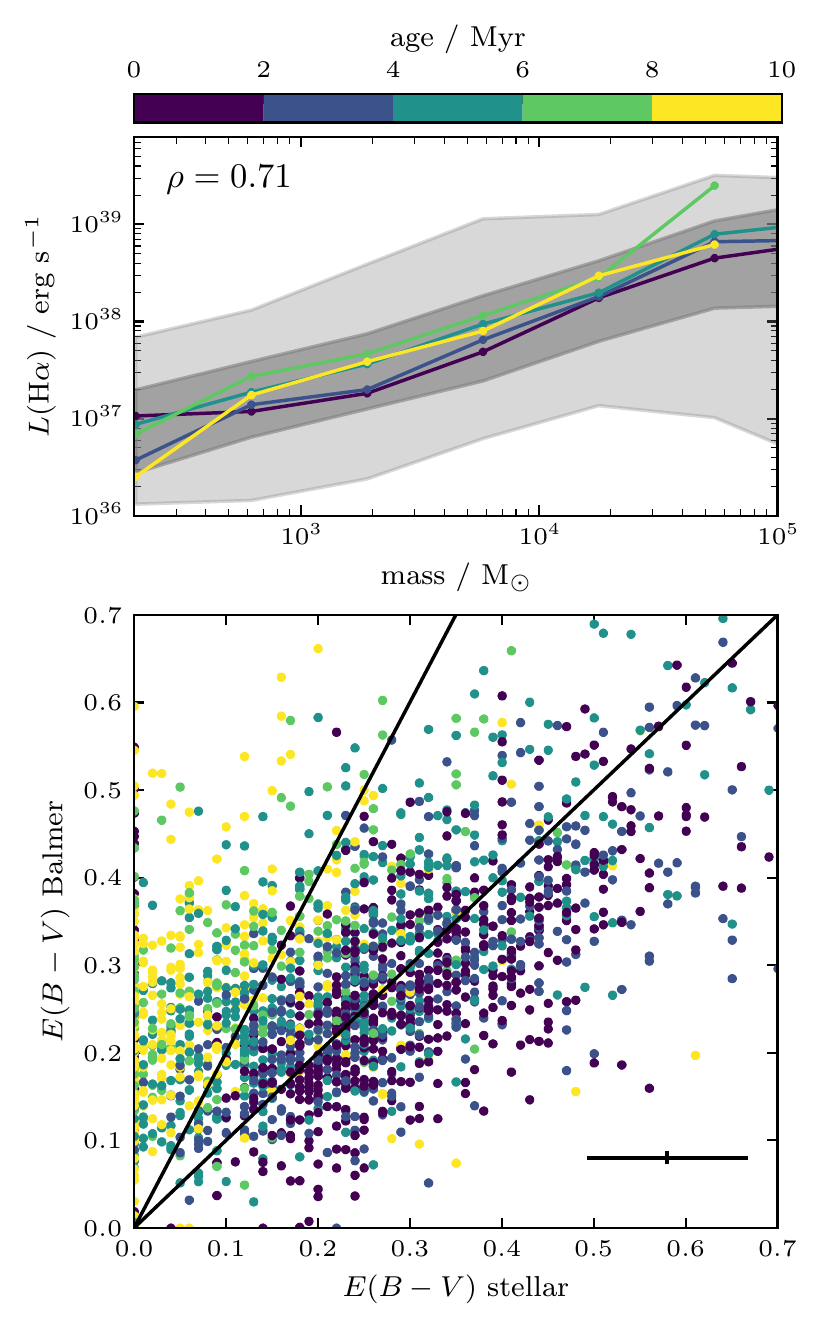}
    \caption{Top: comparison between stellar association mass and extinction corrected $\HA$ luminosity. The 68 and 98 percentile are shaded in grey. Bottom: Comparison between the stellar $E(B-V)$ derived from the SED fit and the one from the Balmer decrement. The black lines indicate one-to-one and 0.44 to 1 correlation \citep{Calzetti+2000}. The average uncertainties are indicated in the bottom right corner. Both panels use the \texttt{contained} objects in the \texttt{one-to-one} sample.}
    \label{fig:matched_catalogues}
\end{figure}

We validate our cross-matched catalogue by comparing physical properties between the nebulae and stellar associations. 
If the associations are the origin of the ionizing radiation, the $\HA$ luminosity of the \HII region is expected to be related to the mass of the association. 
As shown in \cref{fig:matched_catalogues}, we find a strong correlation between the two. 
For this comparison we utilise the sub-sample of fully contained associations (\num{1918} objects).
There is some scatter, as the $\HA$ flux is also dependent on the age of the stellar association and the amount of radiation that escapes the cloud.

We also compare the dust content, as traced by the colour excess $E(B-V)$, derived from the SED fit to the one calculated from the Balmer decrement. 
We find that for young ages, the two methods are in good agreement, while for older associations, the reddening from the Balmer decrement is systematically higher than that from the SED fit. 
This could be due to problems with the SED fit, where young reddened cluster are identified by the fitting routine as old and un-reddened \citep{Hannon+2022}.
The age of some associations ($>\SI{8}{\mega\year}$) raises some doubts if they are the origin of the ionizing radiation. 
For most of those objects, the reddening derived from the Balmer decrement is higher than the one derived from the SED fit, suggesting that the age/reddening degeneracy is not correctly resolved and we therefore exclude those objects from further analysis.
Alternatively, for young clusters the gas and the stars are closely associated, while they decouple over time \citep{Charlot+2000}, which would also result in lower reddening of the older clusters relative to the nebular emission. 
Future inclusion of \textit{JWST} bands and refined age determinations will help us distinguish these scenarios \citep{Whitmore+2023}.

Given the above discussion, we are confident that the stellar associations are clearly linked to the nebulae, enabling us to investigate more deeply evolutionary trends with stellar age. 


%% file: tbl/matched_catalogue_columns.tex
\begin{table}
    \centering
    \caption{Columns of the matched catalogue (\texttt{one-to-one sample}).}
    \begin{tabular}{ll}\toprule
        Column & Description  \\\midrule
        \verb|gal_name| & Name of the Galaxy \\
        \verb|region_ID| & ID of the \HII region \\
        \verb|assoc_ID| & ID of the stellar association \\
        \verb|ra_neb| & Right ascension of the nebulae \\
        \verb|dec_neb| & Declination of the nebulae \\
        \verb|ra_asc| & Right ascension of the stellar association \\
        \verb|dec_asc| & Declination of the stellar association \\ 
        \verb|overlap_neb| & Overlap percentage with stellar association  \\
        \verb|overlap_asc| & Overlap percentage with \HII region \\
        \verb|overlap| & Flag for overlap (isolated, partial, contained) \\
        \verb|environment| & Galactic environment (e.g.~bar,centre,disc)\\
        \verb|neighbors| & Number of neighbouring \HII regions \\  
        \verb|HA6562_lum|$^\dag$ & Extinction corrected $\HA[6562]$ luminosity \\
        \verb|EBV_balmer|$^\dag$ & Colour excess from the Balmer decrement \\
        \verb|EBV_stellar|$^\dag$ & Colour excess from the SED fit \\
        \verb|age|$^\dag$ & Age of the stellar association in $\si{\mega\year}$ \\
        \verb|mass|$^\dag$ & Mass of the stellar association in $\si{\Msun}$  \\
        \verb|{filter}_flux|$^{\dag\ddag}$ & Flux in \textit{HST} band in mJy \\
        \verb|Ha/FUV|$^\dag$ & $\HA$ to $\FUV$ flux ratio (extinction corrected) \\
        \verb|EW_HA|$^\dag$ & $\HA$ equivalent width in $\si{\angstrom}$ \\
        \verb|EW_HA_corr|$^\dag$ & Background corrected $\EW$ in $\si{\angstrom}$ \\
        \verb|logq|$^\dag$ & Ionisation parameter $\log q$\\
        \verb|Delta_met_scal| & Local metallicity offset $\DeltaOH$ \\
        \verb|density|$^\dag$ & Electron density in $\si{\per\cm\cubed}$ \\
        \verb|GMC_sep| & Distance to nearest GMC in $\si{\parsec}$ \\
        \bottomrule\addlinespace
        \multicolumn{2}{l}{${}^\dag$ associated errors are included as \texttt{*\_err}.}\\
        \multicolumn{2}{l}{${}^\ddag$ for \texttt{filter} in \textit{NUV}, \textit{U}, \textit{B}, \textit{V} and \textit{I}.}
    \end{tabular}
    \label{tbl:matched_catalogue_columns}
\end{table}

%% file: tbl/overlap_stat.tex
\begin{table}
\centering
\caption{Level of spatial correlation between \HII regions and associations. We list the total number and percentage of \HII regions (associations) that overlap with $N$ associations (\HII regions) respectively.}
\begin{tabular}{r
                S[table-format=5]
                r
                S[table-format=5]
                r
                }
\toprule
$N$ & \multicolumn{2}{c}{\HII regions} & \multicolumn{2}{c}{Stellar associations} \\
\midrule
0 & 11687 & (57.5\%) & 4596 & (28.2\%) \\
1 & 5673 & (27.9\%) & 9288 & (57.0\%) \\
2 & 1702 & (8.4\%) & 2011 & (12.3\%) \\
3 & 643 & (3.2\%) & 319 & (2.0\%) \\
4 & 291 & (1.4\%) & 57 & (0.3\%) \\
>5 & 345 & (1.7\%) & 24 & (0.1\%) \\
\bottomrule
\end{tabular}
\label{tbl:overlap_statistics}
\end{table}

%% file: tex/04evolutionary_sequence.tex

\section[HII region evolutionary sequence]{\HII region evolutionary sequence}
\label{sec:evolutionary_sequence}

In this section we use the matched catalogue to establish different nebular properties as proxies for the age and then explore how other \HII region properties evolve as the nebula ages.
 
\subsection{Correlations with SED ages}
\label{sec:correlation_sed}

\begin{figure}
    \centering
    \includegraphics[width=1\columnwidth]{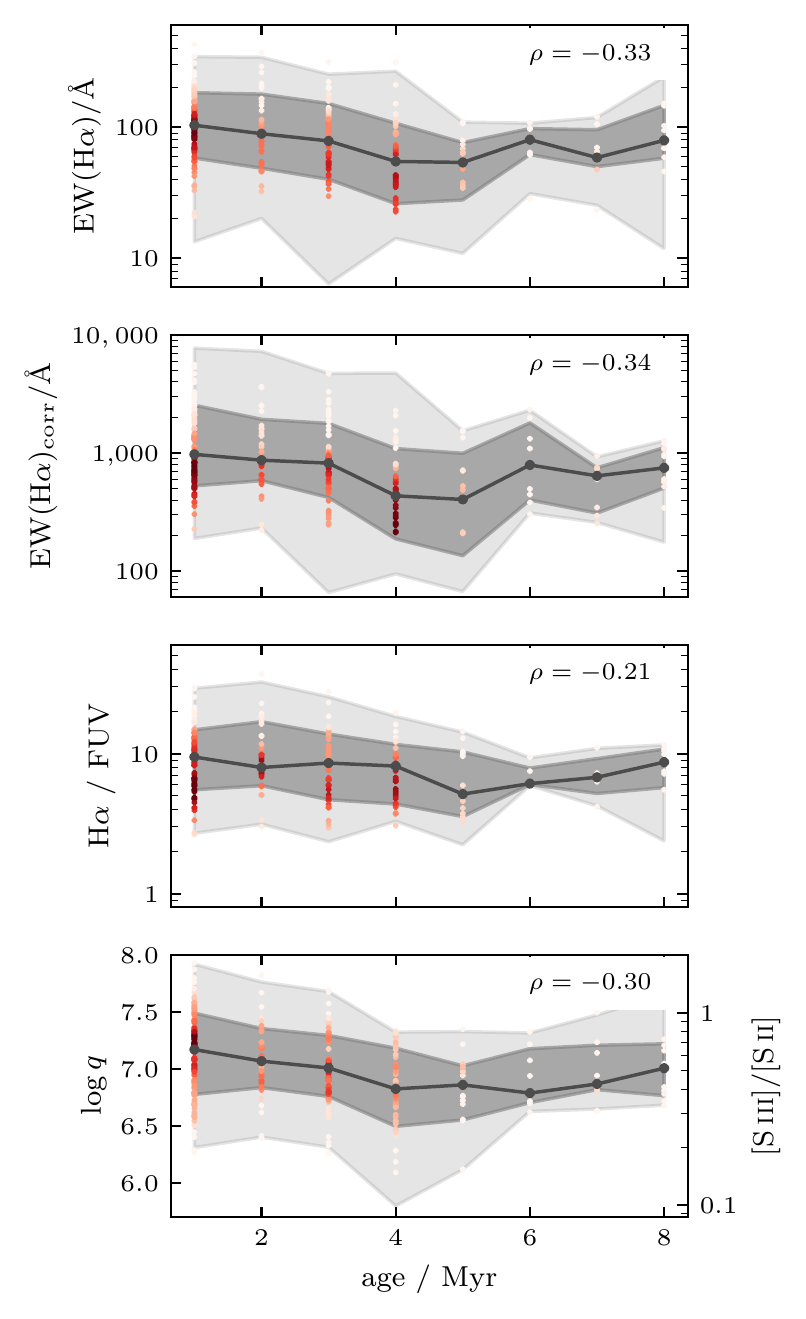}
    \caption{Comparison between stellar associations ages with MUSE \HII region properties. The age is derived from the SED fit and compared to $\EW$, $\EW_\mathrm{corr}$ (the continuum is corrected for background emission), $\HA/\FUV$ and $\log q$. The binned median is shown with a black line and the 68 and 98 percentile are shaded in grey. The colour of the points indicates the density of the points. The Spearman's rank correlation coefficient is shown in the top right corner. The $\EW$, $\EW_\mathrm{corr}$ and $\log q $ have $p$-value$ <\num{e-7}$ and the $\HA/\FUV$ $p$-value$ <\num{e-2}$.}
    \label{fig:age_tracers}
\end{figure}

In \cref{sec:introduction}, we introduced $\EW$, $\HA/\FUV$ and $\log q$ as potential age tracers. 
With the matched catalogue, we can assess how well they trace the age of the stellar population that is powering the nebula. 
Because accurate ages are of utmost importance to this application, we only use the \texttt{robust sample} and also apply further signal-to-noise cuts. 
For the $\HA/\FUV$ and $\log q$ we only use objects with $\StoN\geq 5$ and for the ages we require that they are at least as large as their uncertainties.
This mainly affects $\HA/\FUV$ and choice of a higher $\StoN$ would reduce the sample too much.
For the $\EW$, the largest uncertainty comes from the continuum, which is often indistinguishable from the background. 
We therefore compute $\EW_\mathrm{corr}$ and require that the background subtracted continuum has $\StoN\geq 5$ (this cut is used for the corrected and un-corrected equivalent width). 
In \cref{fig:age_tracers} we compare the age from the SED fit to all four properties.
We observe weak to moderate anti-correlations between the SED ages and all proposed age tracers. 
To test how robust the trends are, we apply a Monte Carlo approach, where we repeatedly sample our data, based on the associated uncertainties and assuming a normal distribution. 
The observed trends are robust and persist, with a scatter around $\SI{10}{\percent}$, except for the $\HA/\FUV$, which shows a larger scatter of $\SI{20}{\percent}$.  

The $\EW$ drops by $\SI{47}{\percent}$ within the first $\SI{4}{\mega\year}$. 
This is caused by the death of the most massive stars and in line with the $\SI{42}{\percent}$ decrease that is predicted by models like \textsc{starburst99} \citep{Leitherer+2014}. 
However, after that, the observed $\EW$ stays roughly constant and even increases slightly towards $\SI{8}{\mega\year}$, compared to the models that predict further decrease to only $\SI{6}{\percent}$ at the same age. 
Also conspicuous is the smaller range of values covered by the observations, compared to the models (see \cref{sec:model_predictions}). 
An easy explanation for this is that we neglected the background. For the stellar continuum around $\HA$, we observe a significant contribution from older stellar populations throughout the galaxy. 
Accounting for this, the $\EW_\mathrm{corr}$ increases by a factor of 10, bringing the measured values closer to the model predictions.
The correlation with the background corrected equivalent width $\EW_\mathrm{corr}$ is slightly stronger than the un-corrected one.
However it is quite challenging to correctly disentangle the contribution of the background.

The flux ratio of $\HA$ to $\FUV$ stays roughly constant for the first $\SI{4}{\mega\year}$, after which it drops to $\sim\kern-2pt\SI{65}{\percent}$.
Again, after that, contrary to the model predictions, the ratio does not further decrease. 
Similar to the equivalent width, the observed values are smaller than the model predictions, although the difference is not quite as large. 
Applying the same reasoning as with the equivalent width to resolve this discrepancy does not work. 
In the case of the $\FUV$, we do not observe a global background, but rather other events of previous nearby star formation, which should not bias our measurements.

Lastly, we look at the ionization parameter. 
For reference we also show $\SIII/\SII$ based on \cref{eq:logq_D91}.
We observe a similar behaviour with an initial drop and a flattening after $\SI{4}{\mega\year}$, corresponding to the decrease in ionizing photon flux and expansion of the nebula.
However it should be noted that the range of ionization parameters that we observe only covers the first few $\si{\mega\year}$ in the models \citep[this could be due to systematic offsets between observations and model grids as highlighted in][]{Mingozzi+2020}.

Given that the uncertainties on the ages are often similar to the ages themselves, we consider a broader statistical analysis by binning our sample by the fiducial age tracers and looking at the median SED ages. 
We find that associations that overlap with an \HII region are significantly younger than those that are isolated. 
When binning our sample based on the $\EW$ we find that the associations with high $\EW$ are younger than their counterparts with low $\EW$ (see \cref{sec:age_hist} for more details). 
A similar but less pronounced trend can also be observed when binning with the $\HA/\FUV$ and $\log q$. 

\begin{figure}
    \centering
    \includegraphics[width=1\columnwidth]{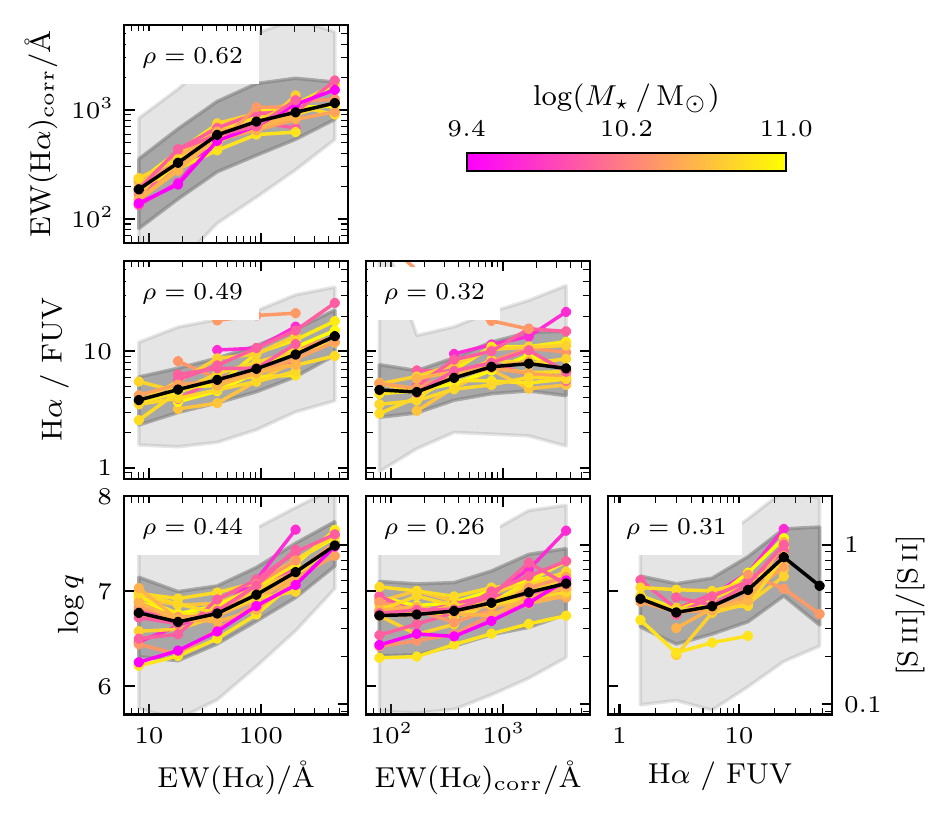}
    \caption{Comparison between the proposed age tracers for the full \HII region catalogue. The nebulae are grouped by their host galaxy, sorted by stellar mass $M_{\star}$, with the median of the entire sample indicated by a black line. The 68 and 98 percentile ranges are shaded in grey.}
    \label{fig:age_tracers_corner}
\end{figure}

Even though the observed decrease with SED ages are weaker than expected, $\EW$, $\EW_\mathrm{corr}$, $\HA/\FUV$ and $\log q$ are all consistent in their evolution. 
In \cref{fig:age_tracers_corner} we use the full \HII region sample of $\sim\kern-2pt\num{20000}$ \HII regions (applying the same signal-to-noise cuts as to the previous sample) to compare the proposed age tracers with each other. 
Across the entire galaxy sample, spanning a wide range of stellar masses, SFRs and metallicities, we find moderate to strong correlations between all properties, consistent with the idea of an evolutionary sequence. 

\subsection{Age trends in the nebular catalogue}\label{sec:trends_nebulae}

\begin{figure*}
\centering
\includegraphics[width=0.9\textwidth]{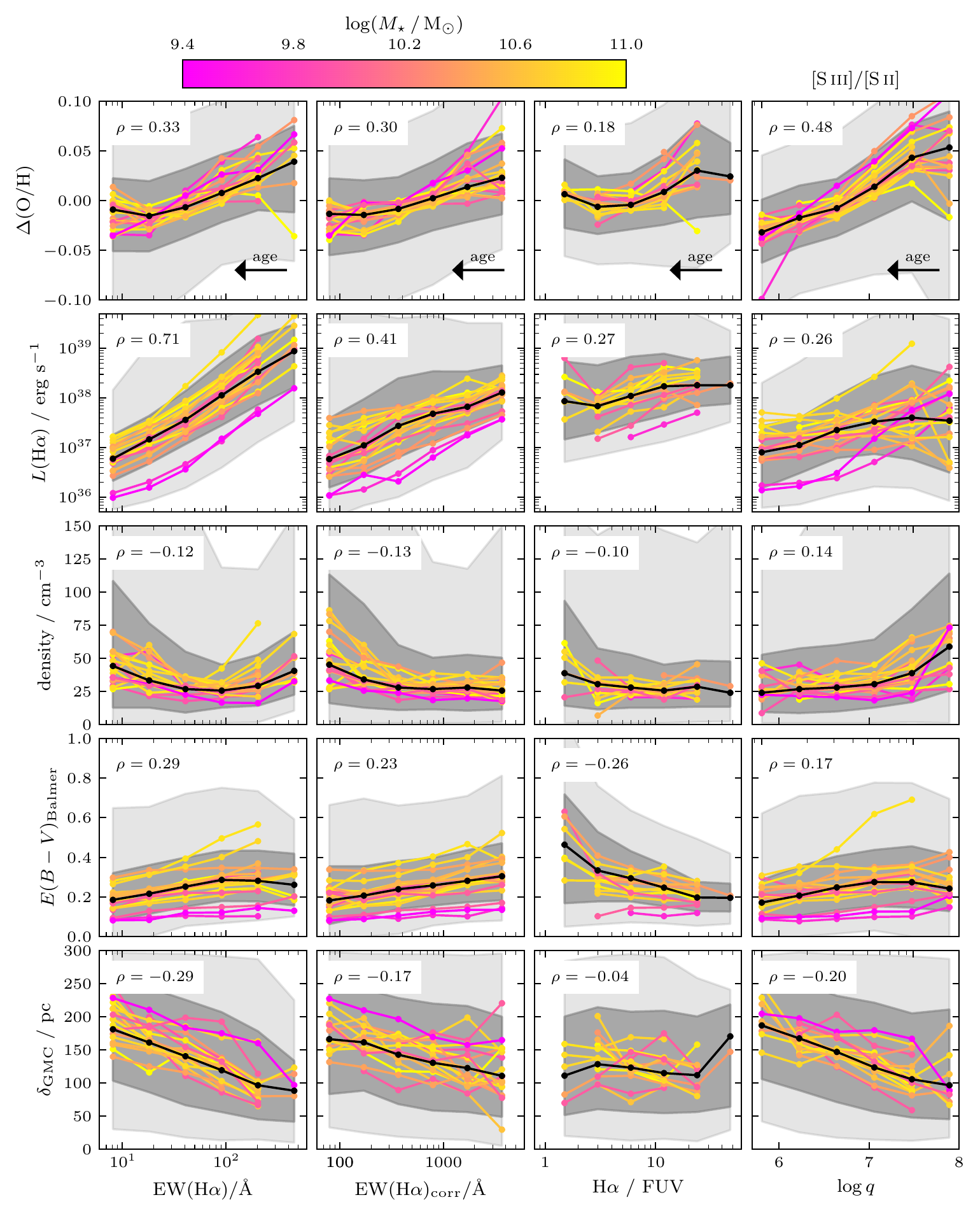}
\caption{Correlations between the physical properties in the \HII region catalogue and the proposed age tracers. We show trends with metallicity offset, $\HA$ luminosity, electron density, colour excess and separation to the nearest GMC. The nebulae are grouped by their host galaxies, sorted by their stellar mass $M_{\star}$, with the average of the entire sample indicated by a black line. The 68 and 98 percentile are shaded in grey.}
\label{fig:trends_with_age}
\end{figure*}

In the previous section, we showed that $\EW$, $\HA/\FUV$ and $\log q$ all exhibit anti-correlations with age, albeit weaker than expected. 
However, for this analysis we were limited to the few hundred \HII regions that contained a massive stellar association, severely limiting our ability to study the evolution of the \HII regions in terms of different ISM properties. 
In this section we use the full catalogue of \num{23736} \HII regions and look for correlations with the proposed age tracers. 
\Cref{fig:trends_with_age} shows trends between $\EW$, $\EW_\mathrm{corr}$, $\HA/\FUV$ and $\log q$ with various nebula properties. 
All trends have $p$-values$<\num{e-6}$, with the exception of $\HA/\FUV$ versus $\delta_\mathrm{GMC}$. 
Following the same procedure as in \cref{sec:correlation_sed}, we find that the trends are robust when varying the properties within their uncertainties.

First, the metallicity variations $\DeltaOH$ are the difference between the local metallicity and the global metallicity gradient. 
We recover the same correlation between $\log q$ and the $\DeltaOH$ that were also found by \citet[][based on the same data]{Kreckel+2019} and \citet{Grasha+2022}. 
The metallicity offset also shows a correlation with $\EW$ and $\HA/\FUV$, indicating that younger regions are more enriched. 
This suggests that the natal environment, not yet dispersed and associated with the youngest regions, is more metal rich.
Over time, this material could mix on larger scales \citep{Kreckel+2020}, decreasing the metallicity offset.

Next, we observe correlations with the $\HA$ luminosity of the \HII regions with our age tracers. The strong correlation with $\EW$ is likely due to the significant contamination of the stellar continuum by (unrelated) older stellar populations. Due to this, the variations could be attributed to the smaller variations in both the $\FUV$ and the stellar continuum compared to $\HA$.

For the electron density of the ionized gas, the majority of our \HII regions fall in the low-density limit, making it difficult to robustly measure a density. 
Only \num{1375} \HII regions are significantly different from the low-density limit which limits our statistical capabilities for individual galaxies.
We find no strong correlations with density, and in most tracers we observe weak anti-correlations, contradicting expectations that newly formed star clusters are still embedded in a dense cloud of gas. 
Focusing on the ionization parameter, we do recover a positive trend, consistent with the simplistic Str\"{o}mgren sphere assumptions from \cref{eq:ionisation_stroemgren}. 

The extinction on the other hand, traced by the colour excess from the Balmer decrement, shows a correlation with $\EW$ and $\log q$, fitting the picture of a young embedded cluster. 
However, by contrast, we observe an anti-correlation with $\HA/\FUV$.
It should be noted that especially for the lower mass galaxies, the trends are very flat.

Finally, we compare with the separation to the nearest \emph{giant molecular cloud} (GMC). 
For this we use the GMC catalogue from \citet{Rosolowsky+2021} and Hughes et al.\ in preparation, based on the PHANGS--ALMA survey \citep{Leroy+2021b}, and cross-match with the \HII region catalogue. 
Given that the typical intercloud and interregion distances in these catalogues are a few hundred $\si{pc}$ (e.g.\ \citealt{Kim+2022}, \citealt{Groves+2023}, Machado et al.\ in preparation) we consider separations of this order to be unphysical but closer clouds may indicate a real separation. 
Therefore we exclude \HII regions with a GMC separation larger than $\SI{300}{\parsec}$ from this analysis. 
We expect the youngest \HII regions to still be closely associated to their birth environment and indeed we find an anti-correlation between separation and age.

For this investigation we presented both the observed and corrected $\EW$. Overall, these two measures show good correlation with each other (\cref{fig:age_tracers_corner}), and similar qualitative trends in \cref{fig:age_tracers}. 
For most properties we find stronger correlations with the un-corrected $\EW$, which could hint at complications with the background subtraction.

\subsection{Understanding the weak correlations with SED age}

%

Although we do see some trends of $\EW$, $\HA/\FUV$ and $\log q$ with age, they are weaker than predicted by models. 
There are a number of reasons why this could be the case. 

First, looking at \cref{fig:catalogue_properties_2D_hist_v2}, we find that roughly half of our sample falls in the youngest $\SI{1}{\mega\year}$ age bin. Given that our median reported uncertainty in the ages is $\SI{1}{\mega\year}$,  the age resolution of the SED fit is not sufficient to probe the evolution of these \HII regions or (realistically) changes that happen within the first \SIrange{1}{2}{\mega\year}. 
Beyond model limitations, another major issue is that we are using only broad bands with limited UV coverage. 
This makes it very difficult to achieve higher precision at very young ages. 
However, the $\sim\kern-2pt\SI{50}{\percent}$ drop in intensity of the nebular age indicators within the first $\SI{4}{\mega\year}$ is broadly consistent with models, and suggests that generally the expected age trends are recovered (see \cref{sec:evolutionary_sequence}).

If we only consider the youngest objects ($\leq \SI{2}{\mega\year}$), we still recover the same trends between $\log q$, $\EW$ and $\HA/\FUV$. 
We note that we also observe the same trends in the older subsample ($> \SI{2}{\mega\year}$), albeit somewhat weaker. 
This suggests that the evolution is not bound to a fixed or absolute timescale, but can vary greatly between clouds. 
In many ways, this is unsurprising, and depending on the environment, the \HII regions can evolve differently.
The timescales over which natal molecular gas clouds are cleared are estimated to be \SIrange{2}{3}{\mega\year} \citep{Kim+2021,Chevance+2022}, suggesting that significant morphological changes in the local environment are occurring. 
Depending on the initial density of the cloud and the ionizing cluster mass, this can lead to large variations in the evolution time scale between different \HII regions \citep{Kruijssen+2019}. 
In this way, we would still expect to see trends between these nebular age tracers, as they are all reflecting the clearing of the cloud, but the absolute timescales (traced by the age of the underlying stellar cluster) would not necessarily agree across different clouds. 

There are also clearly secondary dependencies beyond age, which may or may not significantly impact the trends we explore but are challenging to account for. 
For example, it is likely that different stellar population models would give different results for the age and mass of the stellar associations, and this is not something that has been explored. 
In the case of the ionization parameter, variations will also  arise due to changes in metallicity\citep{Dopita+2014, Kreckel+2019, Grasha+2022} and density \citep{Dopita+2006}, and smear out the trend with age. 
However these relations, particularly with metallicity, are poorly understood from observations or modeling, and it is not even clear if we expect a correlation or anti-correlation \citep[and references therein]{Ji+2022}. 

One key assumption is that we are looking at nebular emission associated with an instantaneous burst of star formation. 
This assumption is not only important for the SED fit, but if the star formation is extended over a larger period of time, it will also be reflected in the observed flux ratios. 
The decrease of both $\HA/\FUV$ and $\EW$ will appear to be slower and much less pronounced, due to the additional underlying stellar continuum emission from the pre-existing older population \citep{Smith+2002,Levesque+2013}. 
Meanwhile, the $\HA$ flux and $\log q$ will largely trace more directly the evolution of the latest burst. 
The exact duration of star formation is actively discussed in the literature, and can vary from very short duration of less than $\SI{1}{\mega\year}$ \citep{Povich+2010} to multiple episodes of star formation across tens of $\si{\mega\year}$ \citep{Ramachandran+2018b}. 
A prominent example here are the two clusters at the centre of 30 Doradus that are thought to have an age spread of a few $\si{\mega\year}$ \citep{Sabbi+2012,Rahner+2018}. 
By studying only a sample where the \HII region and stellar association are matched one-to-one, we have attempted to minimise the impact of blending of different stellar populations.  

One aspect we have neglected is the impact of leaking radiation, which we know must occur as it is largely responsible for the $\si{\kilo\parsec}$ scale ionization of the diffuse ionized gas \citep{Belfiore+2022}.  
Ionizing photons that escape the cloud result in a lower $\HA$ flux measured within the nebula, while the measurement of the $\FUV$ and stellar continuum is mostly unaffected. 
Existing studies measure a wide range of \emph{escape fractions}, ranging from only a few $\si{\percent}$ \citep{Pellegrini+2012} all the way up to \SIrange{60}{70}{\percent} \citep{McLeod+2019,DellaBruna+2021}.
Because the fraction of leaking radiation is likely to change as the cloud ages and dissolves, this can introduce considerable scatter, depending on how much the escape fraction varies between clouds. 
However this process will only decrease the measured $\EW$ or $\HA/\FUV$, and hence the true value should form an upper envelope above the data. 
Escape fractions within the range from \SIrange{40}{80}{\percent} would result in a factor of 5 scatter in $\EW$ at fixed age, roughly consistent with our result.

Overall, a combination of the effects listed above are probably responsible for the observed trends in age being less pronounced than expected. 
However, we emphasise that our observations are still consistent with an evolutionary sequence.
 
\subsection{Recommendations on nebular age tracers}
 
The equivalent width measurement has most commonly been used as an age tracer in the literature, and suffers from relatively few systematics or uncertainties due to extinction corrections and is easy to measure. 
However, the contribution of stellar light from the galaxy disk is problematic. 
The stellar continuum emission arising from old stars in the disk is not physically associated with the nebulae, does not contribute to its ionization, and hence should not compromise its role as an age tracer. 
As suggested in \cref{fig:age_tracers_corner}, the raw $\EW$ appears to correlate strongly with the corrected one $\EW_\mathrm{corr}$, however robust use of this tracer is only possible if we include $\StoN$ cuts on the sample that require an understanding of the background contribution. 

On average, the background contribution makes up $\SI{90}{\percent}$ of the light, such that the uncertainties introduced when converting between observed and galaxy-corrected $\EW$ introduce a lot of scatter. 
This makes this age proxy particularly problematic to use for individual \HII regions. 
As the background correction we currently implement relies heavily on identifying nearby (but not co-spatial) contributions to the continuum stellar light, this could be improved by utilising constraints directly at the location of the \HII regions themselves. 
This should be possible by carrying out physically motivated stellar population fits to the \HII region spectrum itself, separating the contribution of older stars and refining this age tracer.  
However, challenges in identifying robust young stellar templates \citep{Emsellem+2022} make such an analysis beyond the scope of this paper. 

The ionization parameter also shows a lot of promise as an evolutionary tracer, however it is strongly dependent on local physical conditions and is perhaps less robust as a direct age tracer. 
Here we are deriving ionizing parameter based on the $\SIII/\SII$ ratio, which shows a very strong primary correlation and minimal secondary dependencies in photoionization modeling \citep{Kewley+2002}. 
While the wide wavelength separation of the lines means they are susceptible to uncertainties in the extinction correction,  the lines are at relatively red wavelengths and therefore attenuation effects are minimised. 
As $\SIII/\SII$ can be detected in the largest number ($\SI{86}{\percent}$) of our nebulae, it also holds the most promise for exploring  evolutionary trends statistically. 

We deem $\HA/\FUV$ to be the least trustworthy age tracer. 
We note that the trends with other physical properties measured in the nebulae are weaker with $\HA/\FUV$ than with the other two nebula age tracers. 
In addition, there are large uncertainties arising from the extinction correction, as $\HA$ and $\FUV$ are at vastly different wavelengths. 
At these physical scales it is not entirely clear at what stage the reddening in the stars begins to deviate from the reddening measured in the gas (via the Balmer decrement) and we find that our choice of $X$ in ${E(B-V)}_\mathrm{stellar} = X \cdot {E(B-V)}_\mathrm{Balmer}$ can alter some of the observed trends. 
Our comparison with the \textit{HST} SED derived $E(B-V)$ in \cref{fig:matched_catalogues} suggests most of the nebulae are best fit by $X=1$, however this changes at ages roughly older than $\SI{5}{\mega\year}$. 
Another issue is the sensitivity of the observations: we detect $\HA/\FUV$ in less than $\SI{23}{\percent}$ of our \HII regions with a $\StoN\geq5$. 
For these reasons, we are hesitant to recommend $\HA/\FUV$ as an age tracer, but note that it does still encode information about the early evolutionary state of these nebulae. 

%% file: tex/05conclusion.tex
\section{Conclusion}
\label{sec:conclusion}

We combine observations from MUSE, \textit{HST} and \textit{AstroSat} to study \HII regions and their ionizing sources together. 
This enables us to use the SED age from the ionizing stellar associations as clocks and time the evolution of the \HII regions.
\begin{enumerate}[leftmargin=*]
    \item We present a catalogue of \num{4177} \HII regions and stellar associations that are clearly matched to each other. This catalogue is well suited for placing empirical constraints on models.
    \item We find strong correlations between the masses of the stellar associations and the $\HA$ fluxes of the \HII regions as well as the colour excess derived from the Balmer decrement and the SED fit.
    \item We search for age trends and find weak to moderate correlations with the $\EW$, $\HA/\FUV$ and $\log q$.
    \item All three properties show consistent trends among each other, hinting at an evolutionary sequence.
    \item We find similar trends with the raw and corrected $\EW$. However the $\StoN$ cuts that we apply necessitate an understanding of the background in both cases.  
    \item Using our nebular age indicators, we find tentative trends for younger \HII regions to exhibit higher densities, higher reddening and smaller separations to GMCs. Interestingly, we also find strong correlations with local metallicity, with the youngest \HII regions exhibiting locally elevated metallicities.
\end{enumerate}
The catalogue presented in this paper provides a novel statistical base for further investigations of the interaction between newly formed stars and their natal gas cloud. 
For example it can be used to validate models like \citet{Kang+2022} that use emission line diagnostics of \HII regions to predict the properties of the underlying star cluster. 
It also allows us to study the impact of stellar feedback on the ISM and in a follow up paper, we will use this catalogue to measure escape fractions of the \HII regions. 
The addition of PHANGS--\textit{JWST} observations for all 19 targets \citep{Lee+2023} will further serve to refine our view of the early embedded phase of star formation, completing our view of the nebular evolutionary sequence and baryon cycle in these nearby galaxies.

%% file: tex/06appendix.tex

\section{Model predictions}\label{sec:model_predictions}

\begin{figure}
    \centering
    \includegraphics{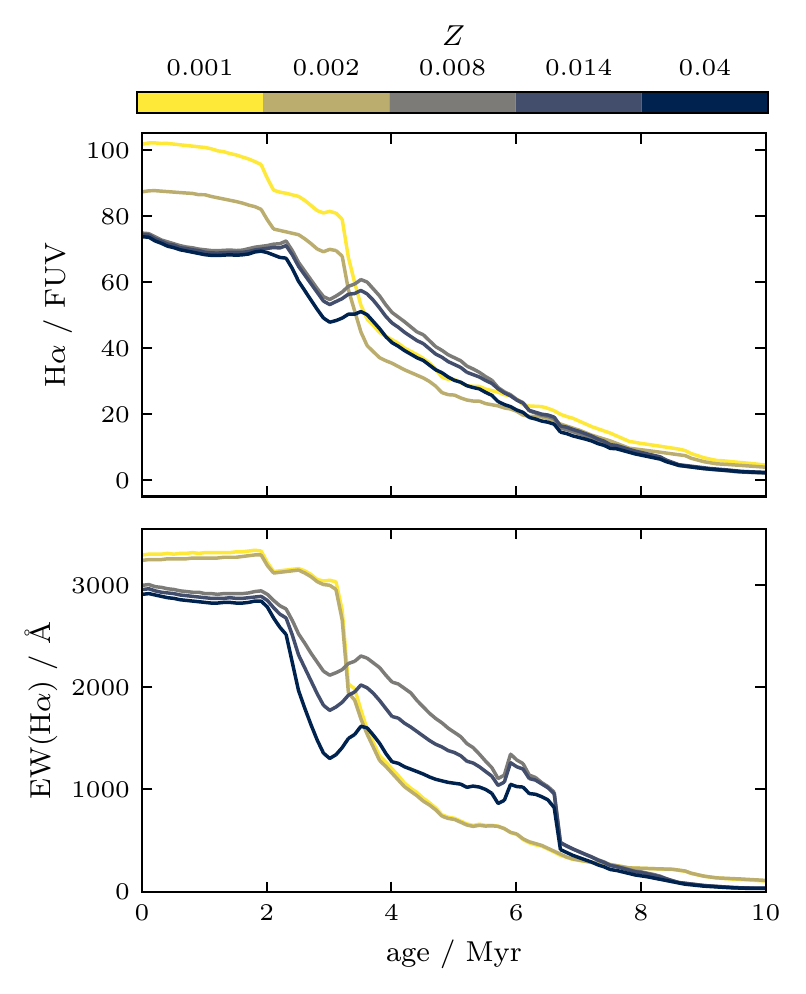}
    \caption{$\HA/\FUV$ flux ratio and $\EW$ as a function of cluster age for different metallicities as predicted by \textsc{starburst99}.}
    \label{fig:age_vs_Ha_over_FUV}
\end{figure}

In \cref{fig:age_vs_Ha_over_FUV} we show the age evolution of $\HA/\FUV$ and $\EW$ as predicted by \textsc{starburst99} \citep{Leitherer+2014}, following a \citet{Kroupa+2001} IMF.
Independent of metallicity, we see a plateau for the first $\sim\kern-2pt\SI{3}{\mega\year}$ after which both the $\HA/\FUV$ and the $\EW$ decrease almost monotonically. 

\section[Background corrected EW]{Background corrected $\mathrm{EW}$}\label{sec:bkg_EW}
We provide the raw and background corrected $\EW$ and $\mathrm{EW}(\HB)$ for the full nebula catalogue from \citet{Groves+2023}. 
The fluxes are measured in the rest frame wavelength intervals listed in \cref{tbl:ew_intervals}, following the procedure from \citet{Westfall+2019}.
To account for the contribution of older stars in the stellar disk, we subtracted the background from an annulus with three times the size of the \HII region. 
For roughly one third of the \HII regions, the background subtraction is complicated by neighbouring \HII regions, and we mask pixels that fall in other nebulae. 
Because the continuum is relatively smooth, this should not be an issue, except for the $\SI{0.5}{\percent}$ \HII regions that are completely surrounded by other nebulae. 
Those objects are hence excluded from the analysis of the $\mathrm{EW}$.

\begin{table}
    \centering
    \caption{Intervals for the equivalent width measurement.}
    \begin{tabular}{lll}\toprule
         & \multicolumn{1}{c}{$\HA$} & \multicolumn{1}{c}{$\HB$} \\\midrule
       Line  & \SIrange{6557.6}{6571.35}{\angstrom} &  \SIrange{4847.9}{4876.6}{\angstrom} \\
       Continuum low  & \SIrange{6483.0}{6513.0}{\angstrom} & \SIrange{4827.9}{4847.9}{\angstrom} \\
       Continuum high  & \SIrange{6623.0}{6653.0}{\angstrom} & \SIrange{4876.6}{4891.6}{\angstrom} \\\bottomrule
    \end{tabular}
    \label{tbl:ew_intervals}
\end{table}

\section{Density and temperature}\label{sec:CrossTemDen}

\begin{figure}
    \centering
    \includegraphics[width=0.8\columnwidth]{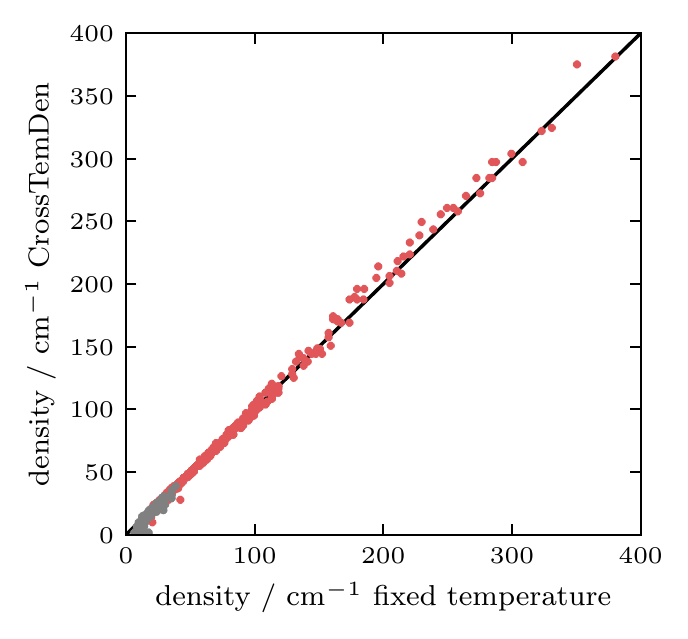}
    \caption{Comparison between the densities derived at a fixed temperature of $\SI{8000}{\kelvin}$ and those derived by fitting density and temperature simultaneously. The grey points are indistinguishable from the low-density limit.}
    \label{fig:getCrossTemDen}
\end{figure}

As discussed in \cref{sec:data}, we assume a fixed temperature of $\SI{8000}{\kelvin}$ to derive the density.
To validate this assumption, we compare the values derived with a fixed temperature versus those fitted simultaneously with the temperature.
Auroral lines are faint, and only \num{840} \HII regions are detected with a $\StoN>10$ in the temperature sensitive $\NII[5754]$ line.
For this sub-sample, we use \textsc{pyneb} \citep{Luridiana+2015} to derive the electron density and the temperature simultaneously from the $\SII[6731/6717]$ and $\NII[5755/6548]$ ratios.
We also derive the density with a fixed temperature of $\SI{8000}{\kelvin}$ and compare the two in \cref{fig:getCrossTemDen}.
As evident in the figure, the small variations in temperature do not affect the measured density significantly. 
While the majority of these \HII regions are significantly different from the low density limit, the sample is only half the size of the full catalogue. 
Due to the large difference in sample size and the small difference in the derived density we opt to use the densities derived with a fixed temperature in our analysis.

\section[Objects in unmatched HII regions]{Objects in unmatched \HII regions}\label{sec:single_stars}

\begin{figure}
    \centering
    \includegraphics{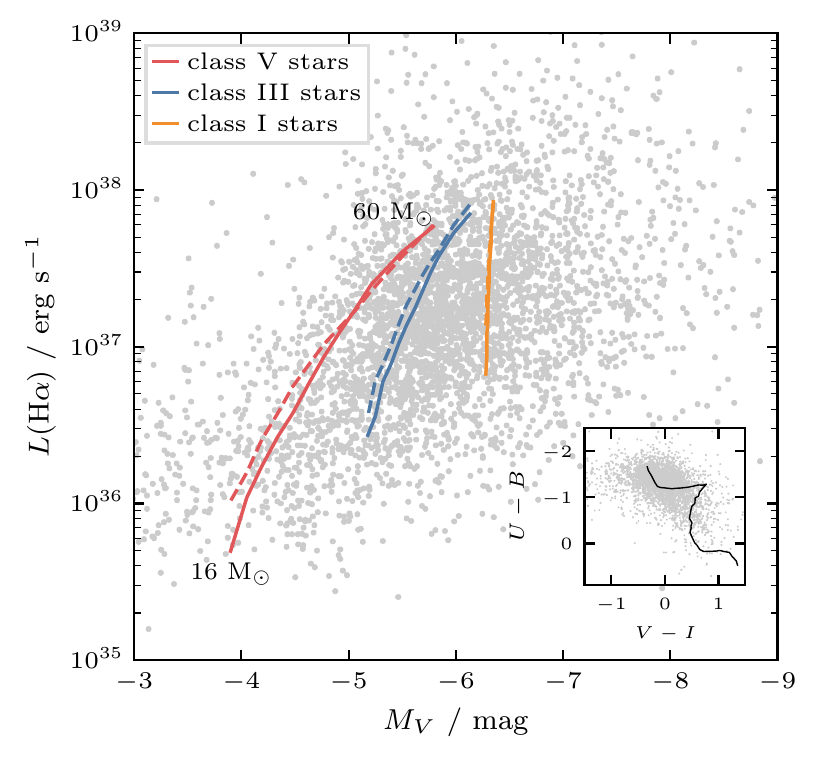}
    \caption{$\HA$ luminosity and absolute \textit{V} band magnitude for the \textsc{dolphot} peaks that are matched to our previously unmatched \HII regions. Overplotted are models for different class O-type stars from \citet{Martins+2005}. The ionizing photon flux is converted to an $\HA$ luminosity via $Q(\HA)=\num{7.31e11} L(\HA)\, \si{\per \second}$ \citep{Kennicutt+1998a}. The solid lines shows their theoretical calibration of $T_\mathrm{eff}$ and the dashed lines the observational calibration. The insert shows a colour-colour diagram of the sample with the track of a single stellar population at solar metallicity for reference \citep{Bruzual+2003}.}
    \label{fig:dolphot_peaks_stellar_models}
\end{figure}

As stated in \cref{sec:unmatched_HII_regions}, almost \num{9000} \HII regions do neither contain a stellar association nor a compact star cluster. 
However, we do find a \textsc{dolphot} peak in $\SI{63.7}{\percent}$ of them.
In \cref{fig:dolphot_peaks_stellar_models}, we compare the $\HA$ luminosity of those unmatched \HII regions with the \textit{V}-band magnitudes of the \textsc{dolphot} peaks and overplot models for single stars from \citet{Martins+2005}. 
We find that most unmatched \HII regions fall in a regime where they could be ionized by a single massive star.

\section{Binned age histograms}\label{sec:age_hist}

\begin{figure*}
    \centering
    \includegraphics[width=\textwidth]{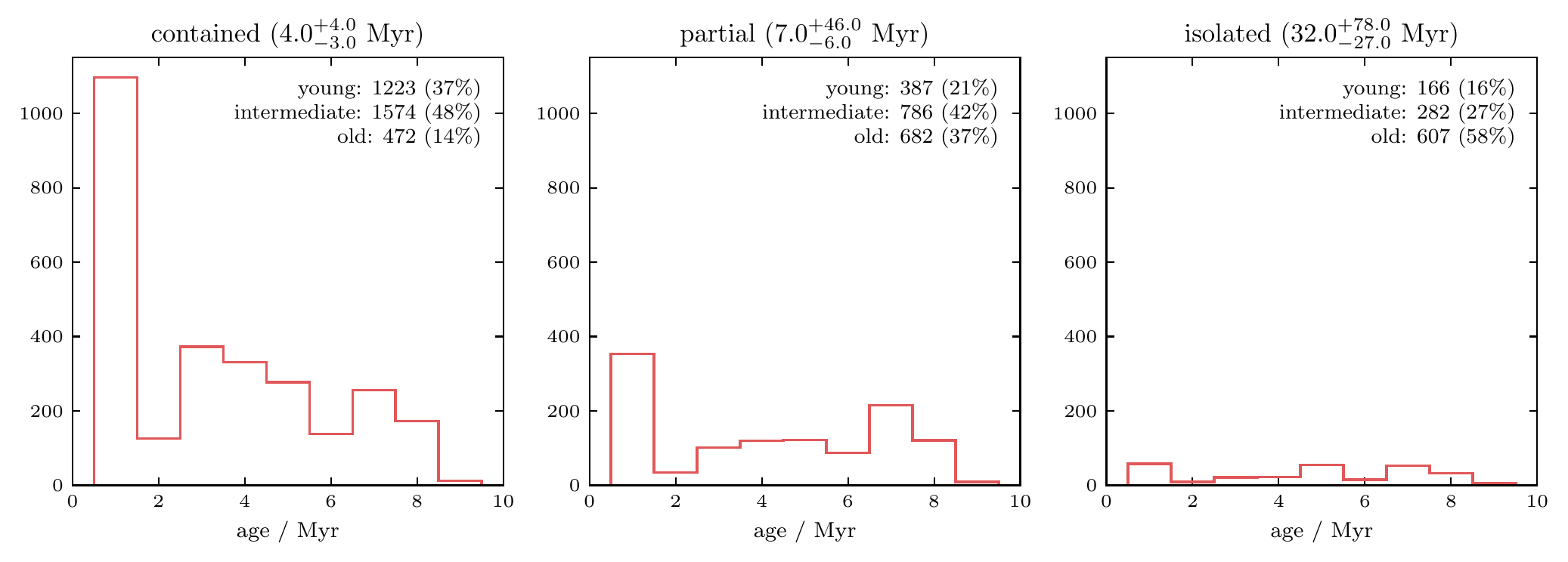}
    \caption{Age distribution of the associations with \texttt{isolated}, \texttt{partial} and \texttt{contained} overlap (only the first $\SI{10}{\mega\year}$ are shown here). For each subsample we show the median age with the uncertainty taken from the $\SI{68}{\percent}$ interval of the data and the number of young ($\leq\SI{2}{\mega\year}$), intermediate (between $\SI{2}{\mega\year}$ and $\SI{10}{\mega\year}$) and old ($>\SI{10}{\mega\year}$) associations. We only include the \num{6027} associations that are more massive than $\SI{e4}{\Msun}$. }
    \label{fig:age_hist}
\end{figure*}

\begin{figure*}
    \centering
    \includegraphics[width=\textwidth]{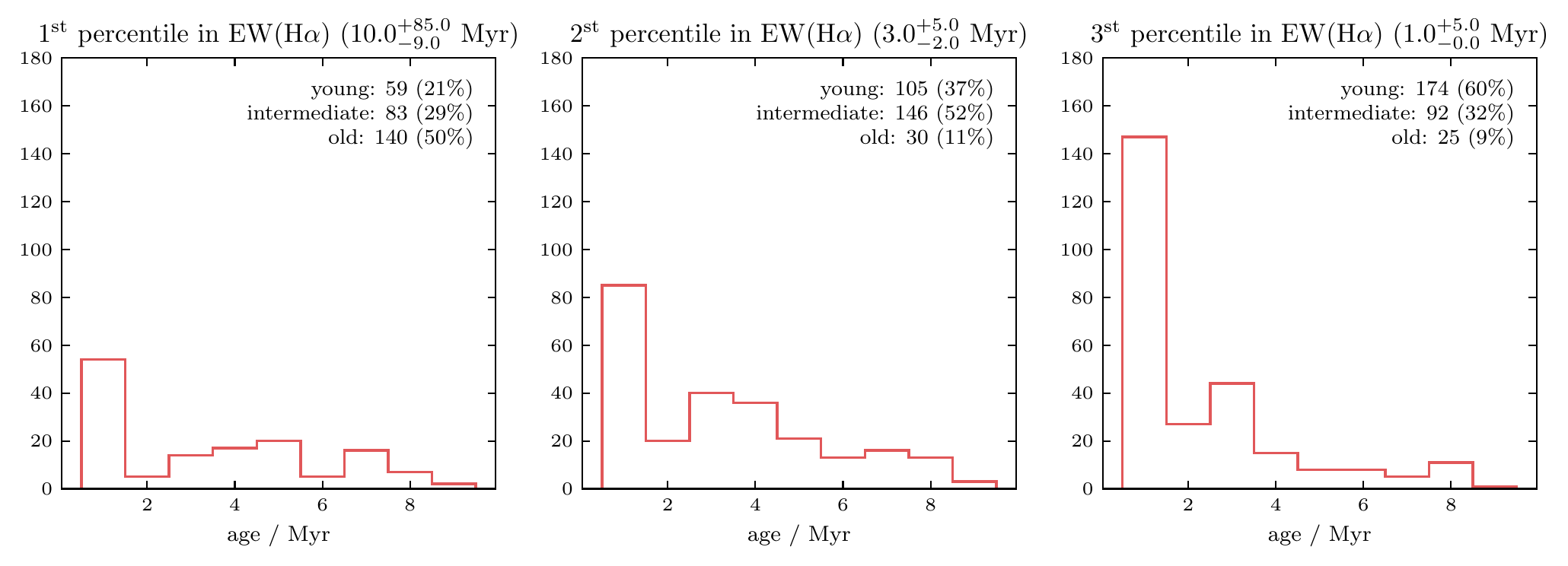}
    \caption{Age distribution, similar to \cref{fig:age_hist}, but based on the $\EW$. We only include associations that are more massive than $\SI{e4}{\Msun}$ and apply the same $\StoN$ cut described in \cref{sec:correlation_sed}, leaving us with \num{854} objects.}
    \label{fig:age_hist_eq_width}
\end{figure*}

\Cref{fig:age_hist,fig:age_hist_eq_width} show the distribution of stellar association ages for different sub-samples. 
In \cref{fig:age_hist} we separate the sample based on the overlap with the \HII regions. We find that associations that overlap with an \HII region are significantly younger than those that are isolated. 
In \cref{fig:age_hist_eq_width} we separate the sample based on the first, second and third percentile in the un-corrected $\EW$. 
The sample with the highest $\EW$ has on average the lowest SED ages.

